\title{April2018-BookCh-LPCopula}
\documentclass[12pt,a4paper]{article}

\usepackage{enumitem}
\usepackage{bold-extra}
\usepackage[comma,authoryear]{natbib}
\usepackage[english]{babel}
\usepackage[utf8]{inputenc}
\usepackage{amsmath,multibib}
\usepackage{amssymb,amsthm}
\usepackage{mathrsfs}
\usepackage{graphicx}
\usepackage{url}
\usepackage{verbatim}
\usepackage{booktabs}
\usepackage{setspace}
\usepackage{multirow}
\usepackage{xcolor,bm}
\usepackage{caption}
\usepackage{setspace}
\usepackage{tcolorbox}
\numberwithin{equation}{section}
\theoremstyle{plain}
\newtheorem{thm}{Theorem}

\newcommand{\bthm}{\begin{thm}}
\newcommand{\ethm}{\end{thm}}
\newcommand{\bpf}{\begin{proof}}
\newcommand{\epf}{\end{proof}}
\theoremstyle{definition}
\newtheorem{defn}{Definition}
\newtheorem{rem}{Remark}[section]
\usepackage[margin=.95in,a4paper]{geometry}
\usepackage[compact]{titlesec}
\renewcommand{\baselinestretch}{1.4}
\usepackage{deep-macro}
\usepackage{pifont}
\newcommand{\cmark}{\ding{51}}%
\newcommand{\xmark}{\ding{55}}%
\usepackage{graphicx}
\usepackage{caption}
\usepackage{subcaption}
\usepackage{mathtools}

\newcommand{\LPsym}{{\rm LPSym}[X \xrightleftharpoons[]{} Y]}
\begin{document}
\begin{center} 
{\Large{\bf Nonparametric Universal Copula Modeling}} \\[.2in] 
Subhadeep Mukhopadhyay and Emanuel Parzen\footnote[2]{{
\scriptsize Shortly after finishing the first draft of this paper, Manny Parzen passed away. Deceased February 6, 2016.}}\\
Final Version\\
\end{center}
\renewcommand{\baselinestretch}{1.3}
\begin{abstract}
\vspace{-.25em}
To handle the ubiquitous problem of ``dependence learning,'' copulas are quickly becoming a pervasive tool across a wide range of data-driven disciplines encompassing neuroscience, finance, econometrics, genomics, social science, machine learning, healthcare and many more.  At the same time, despite their practical value, the empirical methods of `learning copula from data' have been unsystematic with full of case-specific recipes. Taking inspiration from modern LP-nonparametrics \citep{D13a,D13USA,Deep14LP,D12e, Deep17LPMode,deep16LSSD}, this paper presents a modest contribution to the need for a more unified and structured approach of copula modeling that is \textit{simultaneously valid} for arbitrary combinations of continuous and discrete variables. 
\end{abstract}
\vspace{-.4em}
\linespread{1.2}
\begin{center}
{\large {\bf Notation Index}}
\end{center}
\vspace{-1.8em}
\begin{center}
The following summarizes the most commonly used notation of this paper
\end{center}
\vspace{-.8em}
\begin{table}[h]
\centering
\begin{tabular}{p{5cm} p{10cm}}
  \toprule
 Symbol & Description \\
 \midrule
 $X,Y$ & pair of random variables (RVs)\\
 $F_X \equiv F(x;X)$ &  cumulative distribution function (cdf) of $X$\\
 $\Fm(x;F_X)$ &  mid-distribution function of $X$\\
 $Q(u;X)$ & quantile function of $X$\\
 $p(x;X)$ & probability mass function (pmf) for discrete $X$\\
 $f(x;X)$ & probability density function (pdf) for continuous $X$\\
 $F(x,y;X,Y)$ & joint cdf $\Pr(X \le x,Y \le y)$ of $(X,Y)$\\
 $\Cop(u,v;X,Y)$ & copula cdf\\
 $\cop(u,v;X,Y)$ & copula density function\\
 $d(v;Y,Y|X=Q(u;X))$ & conditional comparison density of $Y$ and $Y|F(x;X)=u$\\
 $\LP[j,k;X,Y]$ & $(j,k)$th LP-comean\\
 $\wtF, \tFm$ & The empirical cdf and mid-distribution function\\
 $\Ex[h(X);F_X]$, $\Ex[h(X);\wtF_X]$ & Expectation of $h(X)$ with respect to $F_X$ and $\wtF_X$\\
  \bottomrule
\end{tabular}
\end{table}
\setstretch{1.5}
\section{The Ubiquitous Learning Problem}
How can investors estimate Value-at-Risk (VaR) of a portfolio \citep{embrechts2002}?  How can financial firms assess the joint default probability of groups of risky assets \citep{frey2001copulas}? How can econometricians study the interdependence between family insurance arrangements and health care demand \citep{trivedi2007}? How can neuroscientists delineate the dependence structure among neurons in the brain \citep{berkes2009}? 
How can actuary professionals describe the joint distribution of indemnity payments and loss expenses to calculate the premia \citep{frees1998copula}? How can marketing managers figure out the association between duration of website visits and transaction data to combat low conversion rates \citep{danaher2011}? How can environmental engineers model the dependence structure between hydrologic and climatic variables \citep{aghakouchak2014}?
\vskip.5em
As it turns out, the key statistical challenge to all of these applied multivariate problems lies in developing a method of copula density estimation that is \textit{simultaneously} valid for \texttt{mixed} multivariate data. By `\texttt{mixed},' we mean \textit{any} combination of discrete, continuous, or even ordinal categorical variables. Keeping this end goal in mind, we offer a `one-stop' unifying interface for nonparametric copula modeling.
\section{After 60 Years, Where Do We Stand Now?}
\nocite{sklar1959}
Copula (or connection) functions were introduced in 1959 by Abe Sklar in response to a query of Maurice Fr{\'e}chet. For a pair of random variables $X$ and $Y$, Sklar's theorem states that every joint distribution can be represented as
\beq \label{eq:sklar}
F(x,y;X,Y)=\Cop\big( F_X(x), F_Y(y);X,Y \big),~~~\mathrm{for}~(x,y) \in \cR^2,~~
\eeq 
\vspace{-.5em}
where $\Cop(u,v;X,Y)$ is defined as the joint cumulative distribution function with uniform marginals. For $X$ and $Y$ both continuous (or discrete),  Sklar's copula representation theorem further enables us to decompose the joint density into a product of their marginals times copula density:
\beq \label{eq:copden}
f(x,y;X,Y)=f(x;X) f(y;Y) \cop\big( F_X(x), F_Y(y);X,Y \big), ~~~\mathrm{for}~(x,y) \in \cR^2.~~
\eeq  
Some immediate remarks on copula density function $\cop(u,v;X,Y), 0<u,v<1$:
\begin{itemize}[itemsep=3pt,topsep=1.4pt,leftmargin=10pt]
\vskip.4em
\item Copula density can be interpreted as the ``correction factor'' to convert the independence pdf into the joint pdf. Hence, it acts as the building block of dependence learning.
\item  When $X$ and $Y$ are jointly continuous (or discrete), Copula density can also be expressed as $\cop(u,v;X,Y)=\dep\big(Q(u;X),Q(v;Y);X,Y\big)$, where the dependence function \texttt{dep}, pioneered by \cite{Hoeff40}, is defined as the joint density divided by the product of the marginal densities. A definition of $\cop(u,v;X,Y)$ for general \texttt{mixed} (X,Y) case will be discussed in Section \ref{sec:copdef}.
\item From the statistical modeling perspective, copulas allow us to decouple and separately model the marginal distributions from the dependence structure. 
\end{itemize}
\vskip.2em
For a detailed account of the theoretical properties and probabilistic interpretations of copulas, see the monographs by \cite{schweizer2011, nelsen1999book} and \cite{joe2014book}. In this paper, we shall focus primarily on the statistical modeling principles for fitting copula to the data: 
\begin{quote}
\vspace{-.2em}
\textit{Given a bivariate sample $(X_i,Y_i),$ $i=1,\ldots,n$ how can we estimate the copula density function $\cop(u,v;X,Y)$?}
\vspace{-.3em}
\end{quote}
Typically, the choice of an appropriate statistical estimation algorithm depends on the data-type information (e.g., discrete or continuous) of $X$ and $Y$:

\begin{itemize}[itemsep=3pt,topsep=1.4pt,leftmargin=10pt]
\vskip.45em
\item $X,Y$ both continuous:  A wide variety of parametric copulas exist \citep[Table 4.1]{nelsen1999book} for jointly continuous margins. Among them,  elliptical copulas (including Gaussian and Student-t families), the Archimedean class (including the Clayton, and Frank families), and the extreme-value copulas (including  Gumbel and Galambos) are the most commonly used types. Given this bewilderingly large collection, practitioners often face stiff challenges when choosing an appropriate one for their empirical problem. In addition, these parametric classes of copula models are notoriously less flexible when it comes to capturing real-world complex dependency structures. As a remedy, several nonparametric kernel-based procedures have been developed in recent times; see, e.g., \cite{gijbels1990,chen2007} and \cite{geenens2017}.

\item $X,Y$ both discrete: Copula estimation is far more challenging for discrete margins that include binary, ordinal categorical, and count data. Some promising parametric approaches are discussed in \cite{panagiotelis2012} and \cite{smith2012}. On the other hand, the current state of nonparametric copula estimation for discrete marginals is really grim, and still an open issue; see for example \cite*{genest2007}, which warns practitioners against ``naive'' rank-based estimators. Quite surprisingly, however, the paper offers no practical solution for the problems. Section \ref{sec:ber} will discuss these issues in details. 
\item $X$ discrete, $Y$ continuous: The mixed discrete and continuous case is probably the most challenging regime for copula modeling. Existing parametric attempts \citep{craiu2012mixed,marbac2017model,zilko2016copula} are indirect and forceful. They are based on continuous latent variables, which capture only limited dependence structure, due to their strong reliance on latent Gaussian or parametric copula assumptions. And they are often, computationally, extremely costly. Whereas, perhaps not surprisingly, the nonparametric tools and methods are very much in their nascent stage; see e.g.,  \cite{racine2015mixed} and \cite{nagler2017generic}, which struggle to extend kernel density estimates for mixed data by introducing quasi-inverse or by adding noise (jittering) merely to make the data forcibly continuous-like. Consequently, this class of randomized methods not only makes the computation clumsy but also introduces artificial variability, leading to potential irreproducible findings. 
\end{itemize}
\vskip.2em
{\bf The Motivating Question}. It is evident that current best practices are overly specialized and suffer from ``narrow intelligence.'' They are carefully constructed for each combination of data-type on a case-by-case basis. Thus, a natural question would be:
\begin{quote}
\vspace{-.3em}
\textit{Instead of developing methods in a `compartmentalized' manner for each data-type separately, can we construct a more organized and unified copula-learning algorithm that will work for any combination of $(X,Y)$?}
\vspace{-.3em}
\end{quote}
A solution to this puzzle can have two major impacts on theory and practice of copula modeling. First, it can radically \textit{simplify} the practice by automating the model-building process. This auto-adaptable property of our nonparametric algorithm  (which adapts to different varieties of data automatically) could be especially useful for constructing higher-dimensional copulas according
to a sequence of trees (vine copula; \cite{joe1996}), by using bivariate copulas as building blocks. Second, it can provide, for the first time, a \textit{unified} and holistic understanding of the `science' behind copula learning. It is somewhat remarkable that this question has not been posed in the copula literature before.
\vskip.15em
In the sequel, we introduce a modern formulation of this problem. Our theory is based on a new nonparametric representation technique to provide a systematic and automatic learning pipeline for copula density function.
\section{A Taste of Next-Generation Copula Learning}
Before going into the main theory, here we present a glimpse of where we are headed in our effort to develop a practical and flexible copula model, \texttt{LPCopula}(X,Y)---a model that is exceptionally simple to use, due to its ability to adapt automatically to the underlying data type. We feel this is vital to ensure the safety of the proposed technology in the hands of applied researchers who are not trained in the esoteric theory of nonparametric data science.

\begin{figure}[t]
  \centering
  \includegraphics[width=.46\linewidth,keepaspectratio,trim=2cm 1.25cm 1.75cm 1.75cm]{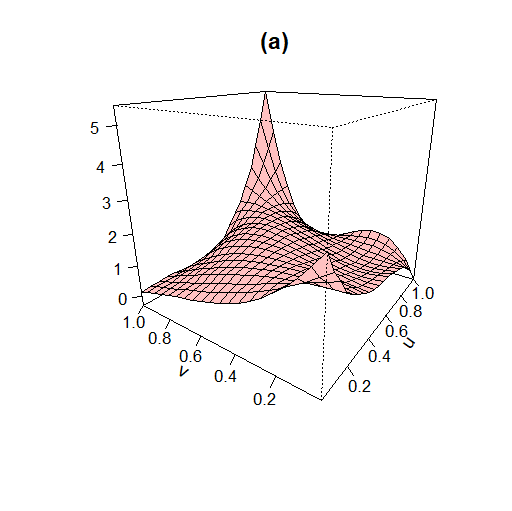}~~~~
\includegraphics[width=.46\linewidth,keepaspectratio,trim=1.75cm 1.25cm 2cm 1.75cm]{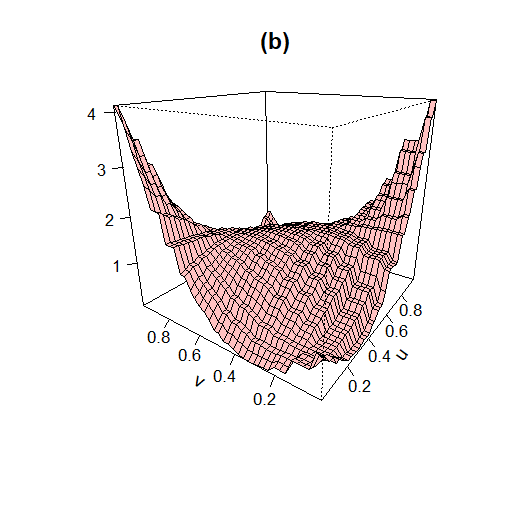} \\[-.25em]
  \includegraphics[width=.46\linewidth,keepaspectratio,trim=2cm 1.75cm 1.75cm 1.25cm]{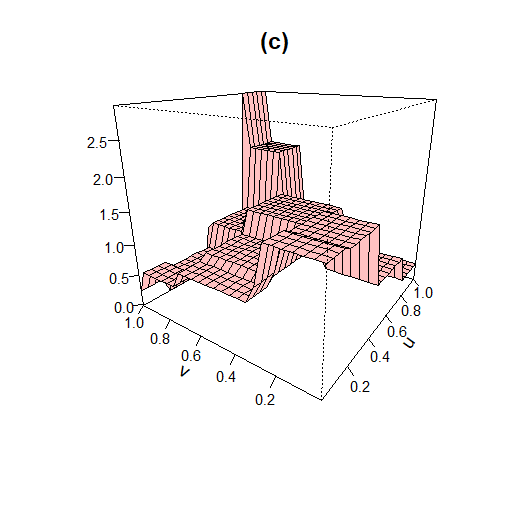}~~~~
  \includegraphics[width=.46\linewidth,keepaspectratio,trim=1.75cm 1.75cm 2cm 1.25cm]{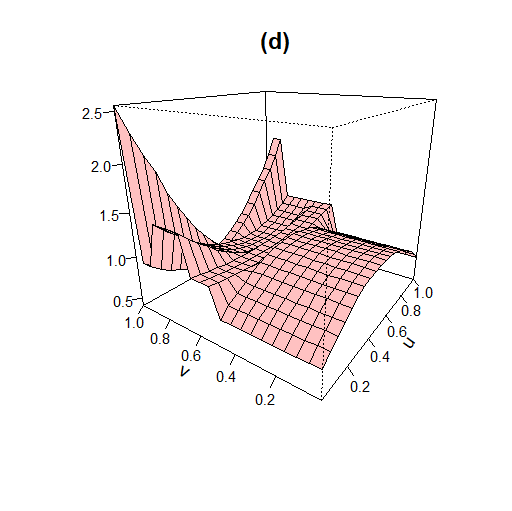}
\vspace{-.35em}
\caption{LP-nonparametric copula density estimate for (a): Insurance data (Loss vs ALAE), (b): Zelterman data (salary vs years since first degree), (c) Fisher's  data (eye vs hair color), and (d): Wooldridge's data (wages vs number of dependents).}
\label{fig:copulaintro}
\vspace{-.5em}
\end{figure}

Using the methodology detailed in the next section, we estimate copula density function for four real data sets, covering the full spectrum of discrete, continuous, and categorical variables. The results are shown in Fig \ref{fig:copulaintro}.
\vskip.25em
{\bf Example 1}. \textit{Both continuous}: This is the classic Loss-ALAE dataset from \cite{frees1998copula}, concerning indemnity payment (LOSS) and allocated loss adjustment expense (ALAE) from $n=1,500$ insurance claims. The estimated copula based on uncensored observations (1,466 in total) is shown in Fig. \ref{fig:copulaintro}(a). 
\vskip.25em

{\bf Example 2}. \textit{Both discrete count}: 
This data \cite[Table 1]{zelterman1987} reports a survey of $n=129$ women employed as mathematicians or statisticians based on their monthly salary and years since their bachelor's degree. The summarized data can be represented as a large, sparse contingency table. It is worthy of remark that estimating copula from this kind of data is known to be challenging. Fig. \ref{fig:copulaintro}(b) displays the staircase $\hcop(u,v;X,Y)$.  
\vskip.25em
{\bf Example 3}. \textit{Both ordinal}: The dataset is based on the cross-classification of people in Caithness, Scotland, by eye color (blue, light, medium, dark) and hair color (fair, red, medium, dark, black), first analyzed by \cite{fisher1940}. This is historically important data, and is believed to be the first example of measuring association in contingency tables. Fig. \ref{fig:copulaintro}(c) displays the estimated piecewise-constant checkerboard copula function.
\vskip.25em

{\bf Example 4}. \textit{Mixed variables}: The data is cross-section wage data consisting of a random sample taken from the U.S. Current Population Survey for the year 1976 \citep{wooldridge2003}. The copula density in Fig. \ref{fig:copulaintro}(d) models the dependence between wages (continuous) and number of dependants (discrete).
\vskip.5em
In contrast to the traditional copula-estimation techniques, the method presented here works across the board without a single modification or tuning in its architecture. It thereby achieves the important goal of automated learning in the mixed-data environment. A detailed description of the model specification, estimation, and exploratory analysis will be discussed in the ensuing sections.
\section{United LP-Nonparametric Methods} \label{sec:LPtheory}
This section provides a fundamentally new copula-representation theory, along with nonparametric estimation strategies that work for \texttt{mixed} data. 

\subsection{Generalized Copula and Conditional Comparison Density} \label{sec:copdef}
United statistical theory aims to unify methods for continuous and discrete random variables. Thus it is important to introduce a bona fide definition of copula density that is valid for \texttt{mixed}(X,Y). This will be achieved via a new representation of Bayes' theorem.
\vskip.4em
We start by noting that when $X$ and $Y$ are both continuous, or both discrete, their joint probability is described by joint probability density $f(x, y; X,Y )$ or by joint probability mass function $p(x, y; X, Y )$. When $Y$ is discrete and $X$ is continuous, the joint probability is described by either side of identity
\beq \label{eq:preBayes}
\Pr[Y = y \mid X = x]f(x; X) = f(x; X \mid Y = y) \Pr[Y = y],
\eeq
which we call the \textit{Pre-Bayes theorem}. Recall that the Bayes' rule is given by $\Pr[Y = y \mid X = x]/\Pr[Y = y] = f(x; X \mid Y = y)/f(x; X)$.

\begin{defn}[Comparison Density]
Define comparison density between $X$ and $Y$ continuous as
\[d(u;X,Y)=\dfrac{f(Q(u;X);X)}{f(Q(u;X);Y)},~~0<u<1\]
and for $X$ and $Y$ discrete, defined in terms of probability mass function:
\[d(u;X,Y)=\dfrac{p(Q(u;X);X)}{p(Q(u;X);Y)},~~0<u<1.\]
\end{defn}
We can now represent Bayes' rule, after the quantile transformation $x = Q(u; X), y = Q(v; Y )$ using conditional comparison density notation:
\beq \label{eq:Brule}
\text{{\bf Bayes' Rule}}:~ d\big(v;Y, Y|X=Q(u;X)\big) = d\big(u;X,X|Y= Q(v; Y)\big),\, 0<u,v<1.
\eeq
Note that For continuous margins case, copula admits the following conditional distribution-based representation:
\beq \label{eq:copC}
\cop\big( F_X(x), F_Y(y);X,Y \big)\,=\,\dfrac{f(y;Y|X=x)}{f(y;Y)} \,=\,\dfrac{f(x;X|Y=y)}{f(x;X)}~~~(x,y) \in \cR^2.
\eeq
We now generalize this for mixed (X,Y) case in the following definition. 
\begin{defn}[Generalized Copula Density]
The key is to recognize that, for $Y$ discrete and $X$ continuous,  we can define \textit{generalized} copula density through conditional comparison density as
\beq \label{eq:copG}
\cop(u, v; X,Y)\,=\,d\big(v;Y, Y|X=Q(u;X)\big)\,=\, d\big(u;X,X|Y= Q(v; Y)\big).\eeq
\end{defn}
\vspace{-.4em}
Here Bayes' theorem \eqref{eq:Brule} asserts the equality of two comparison densities whose value is defined to be copula density. To the best of our knowledge, this is the first rigorous general-purpose definition of copula that is valid for \textit{arbitrary} random variables. 

\begin{defn}[Sequential Multivariate Mixed Copula]
The next result generalizes formula \eqref{eq:copG} to multivariate case $X_1,\ldots,X_d$ via successive conditional comparison densities: 
\beq 
\cop(u_1,\ldots,u_d;X_1,\ldots,X_d) = \text{{\small $\prod_{j=2}^d d\big( u_j; X_j,X_j \mid X_1=Q(u_1;X_1),\ldots, X_{j-1}=Q(u_{j-1};X_{j-1}) \big)$.}}
\eeq
For that reason we call this result ``sequential'' multivariate copula decomposition, valid for mixed discrete and continuous variables. 
\end{defn}

\subsection{Notation and Background}

{\bf Basics 1. Mid-Distribution Transform}. The mid-distribution function of a random variable $X$ is defined as $\Fm(x;F_X)=F_X(x)-\frac{1}{2}p(x;F_X)$ where $p(x; F_X)$ is probability mass function. The $\Fm(X;F_X)$ has mean $\Ex[\Fm(X;F_X)]=.5$ and  variance $\Var[\Fm(X;F_X)]=\frac{1}{12}\big( 1- \sum_x p^3(x;F_X) \big)$. The empirical cdf will be denoted by $\wtF$.
\vskip.5em
{\bf Basics 2. Pseudo-Observations Construction}. Our nonparametric copula approach aims to model the distribution of $\{U, V\} =\big\{\Fm(X;F_X), \Fm(Y;F_Y)\big\}$.  A major obstacle in applying and estimating copula densities is that the marginal of $X$ and $Y$ are unknown. Our approach starts with the mid-distribution function of the sample marginal distribution functions of $X$ and $Y$ to transform observed $(X,Y)$ to $(\wtU,\wtV)$
\beq \label{eq:pob}
\wtU_i = \Fm\big(x_i;\wtF_X\big),~\,\text{and}~\, \wtV_i = \Fm\big(y_i;\wtF_Y\big),~~\text{for}~i=1,\ldots,n.\eeq
It is important to keep in mind that the empirical cdfs $\wtF_X(x)$ and $\wtF_Y(y)$ are discrete, \textit{irrespective} of the data-type of the original $X$ and $Y$. 
\begin{rem}
We recommend displaying the original as well as the copula scatter plots: $(X,Y)$ and $(\wtU,\wtV)$ for a better understanding of the relationship.
\end{rem}
\begin{rem}
Note that our definition of pseudo-samples \textit{avoids} the questionable practice of random tie-breaking using jittering, which is known to mask the real pattern in the data by injecting fake variation and randomness. 
\vspace{-.65em}
\end{rem}


\subsection{Theory and Approximation Methods}
We present our nonparametric theory and methodology in a `programmatic' style--by gradually introducing the essential tools and building blocks that is translatable into an algorithm. 

\begin{figure}[t]
\centering
 \includegraphics[width=.23\textwidth,trim=1cm 1cm 1cm 1cm]{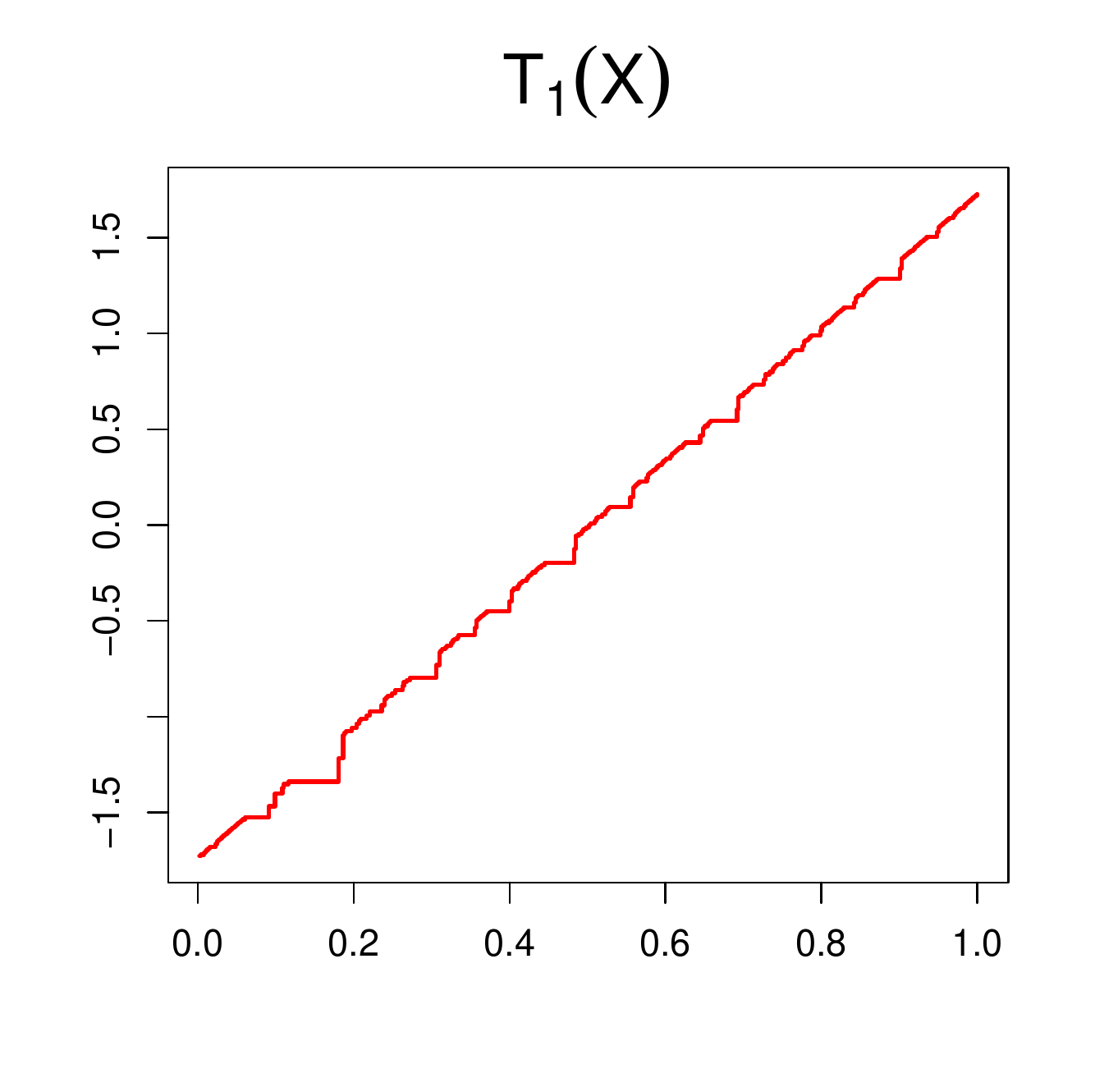}~\,
  \includegraphics[width=.23\textwidth,trim=1cm 1cm 1cm 1cm]{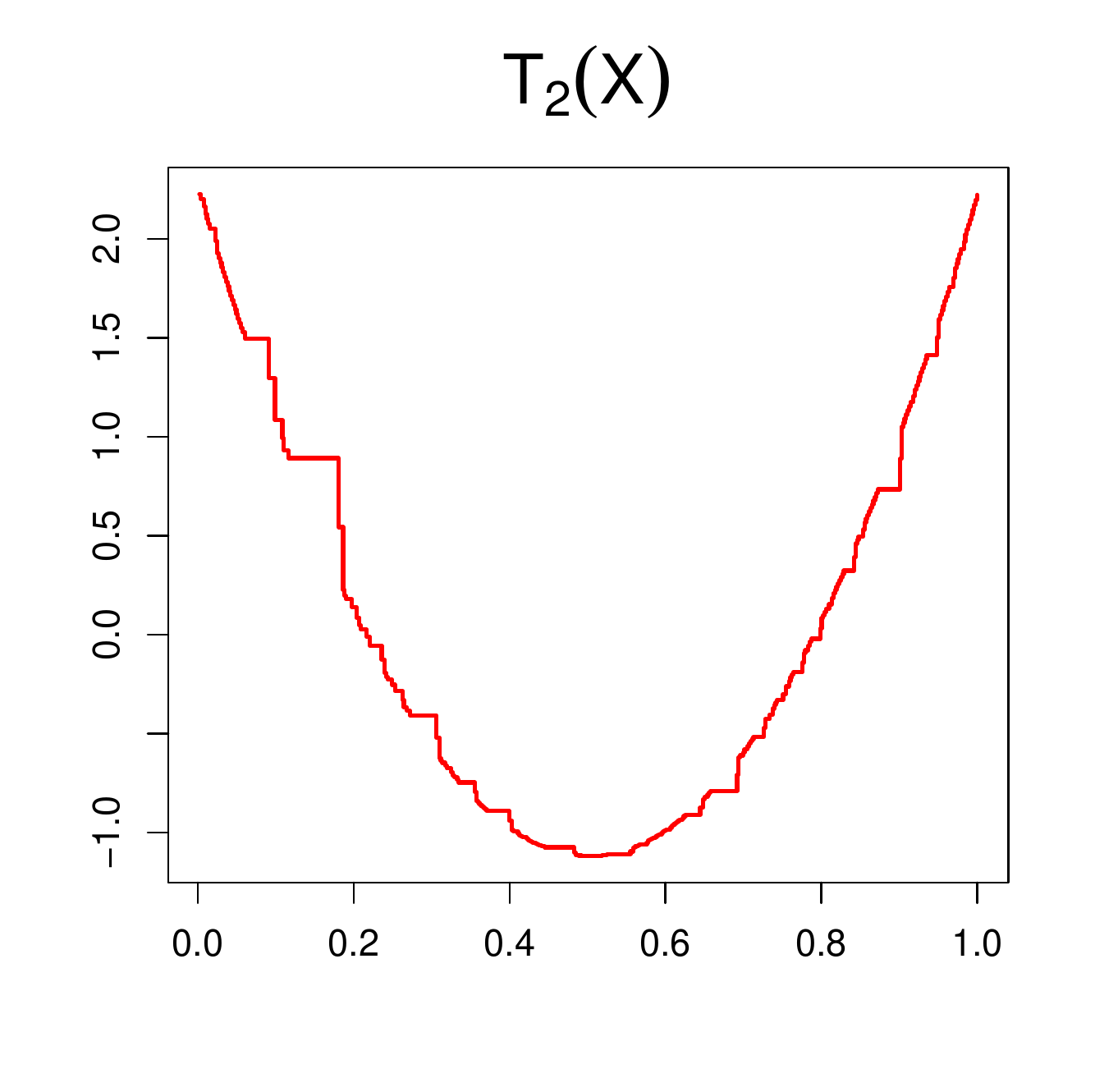}~\,
   \includegraphics[width=.23\textwidth,trim=1cm 1cm 1cm 1cm]{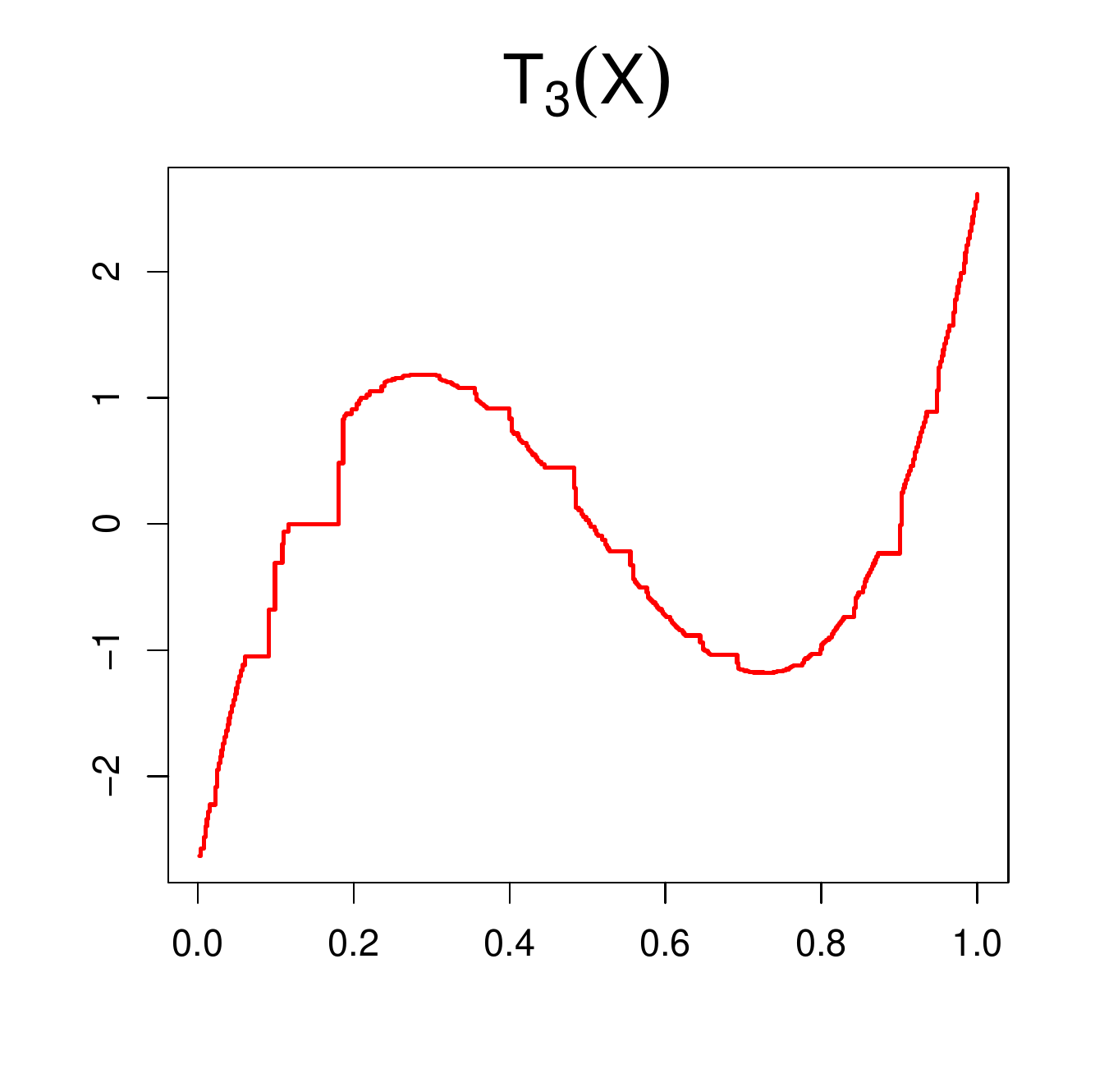}~\,
    \includegraphics[width=.23\textwidth,trim=1cm 1cm 1cm 1cm]{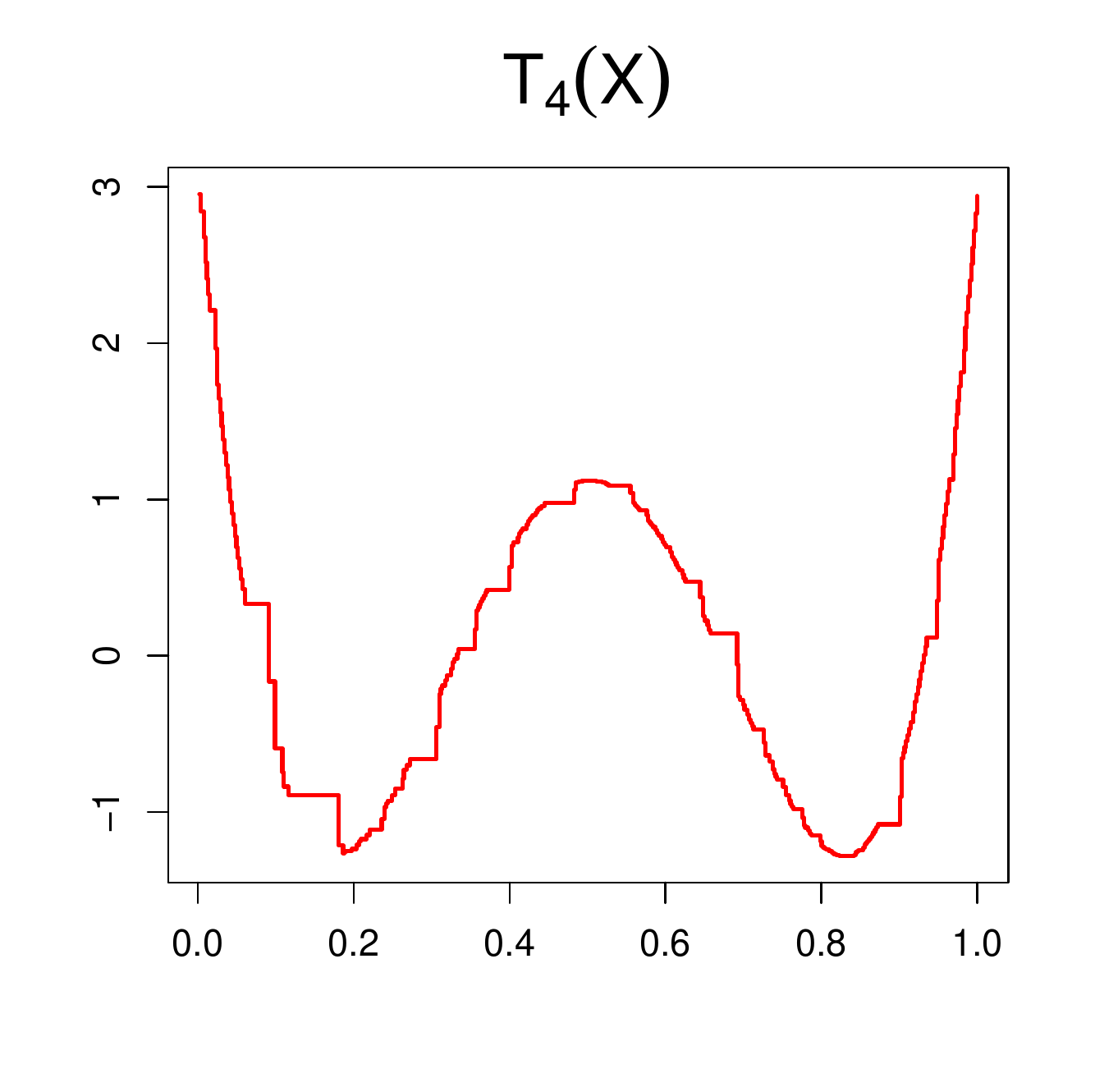}\\[1em]
 \includegraphics[width=.23\textwidth,trim=1cm 1cm 1cm 1cm]{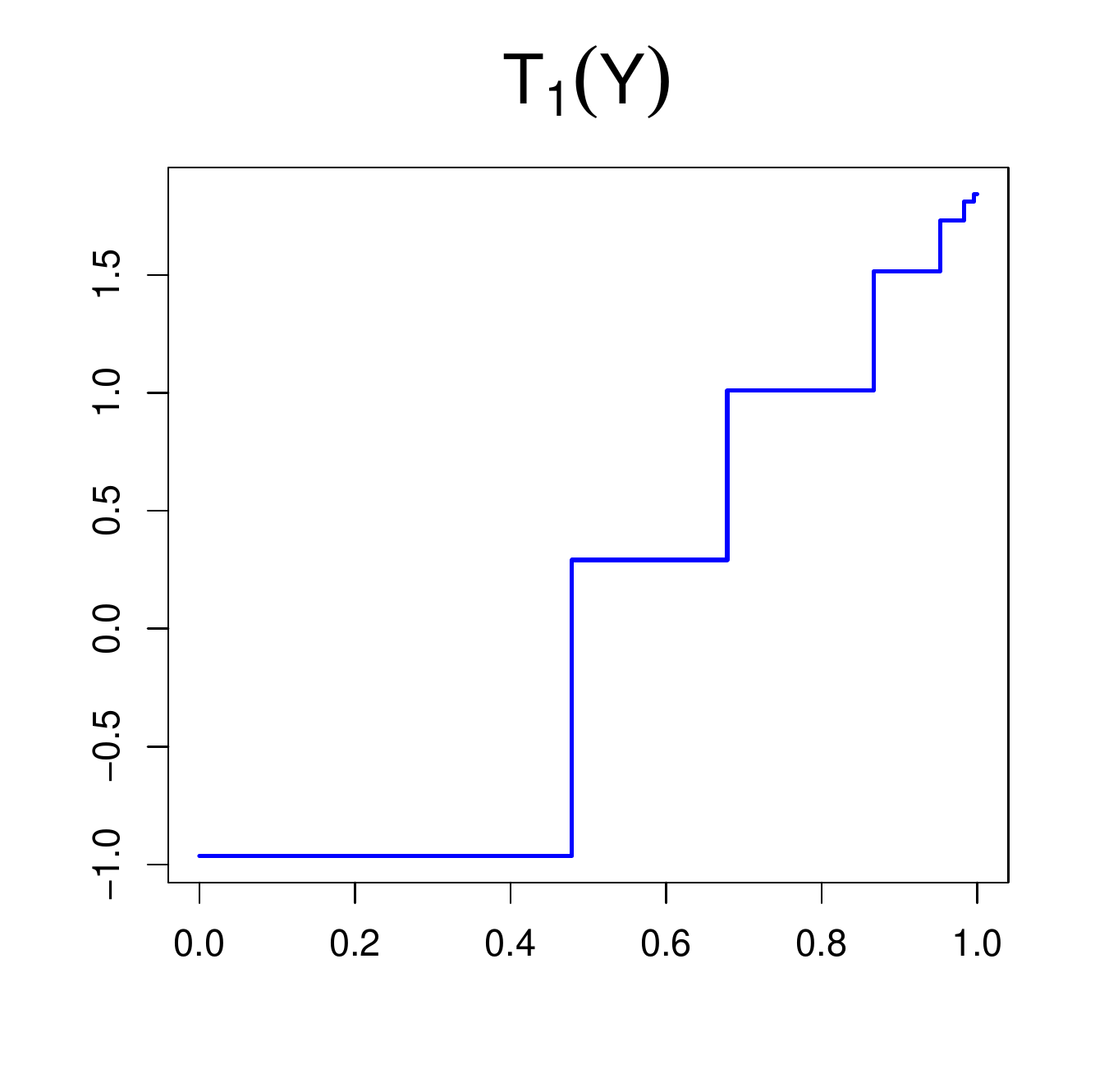}~\,
  \includegraphics[width=.23\textwidth,trim=1cm 1cm 1cm 1cm]{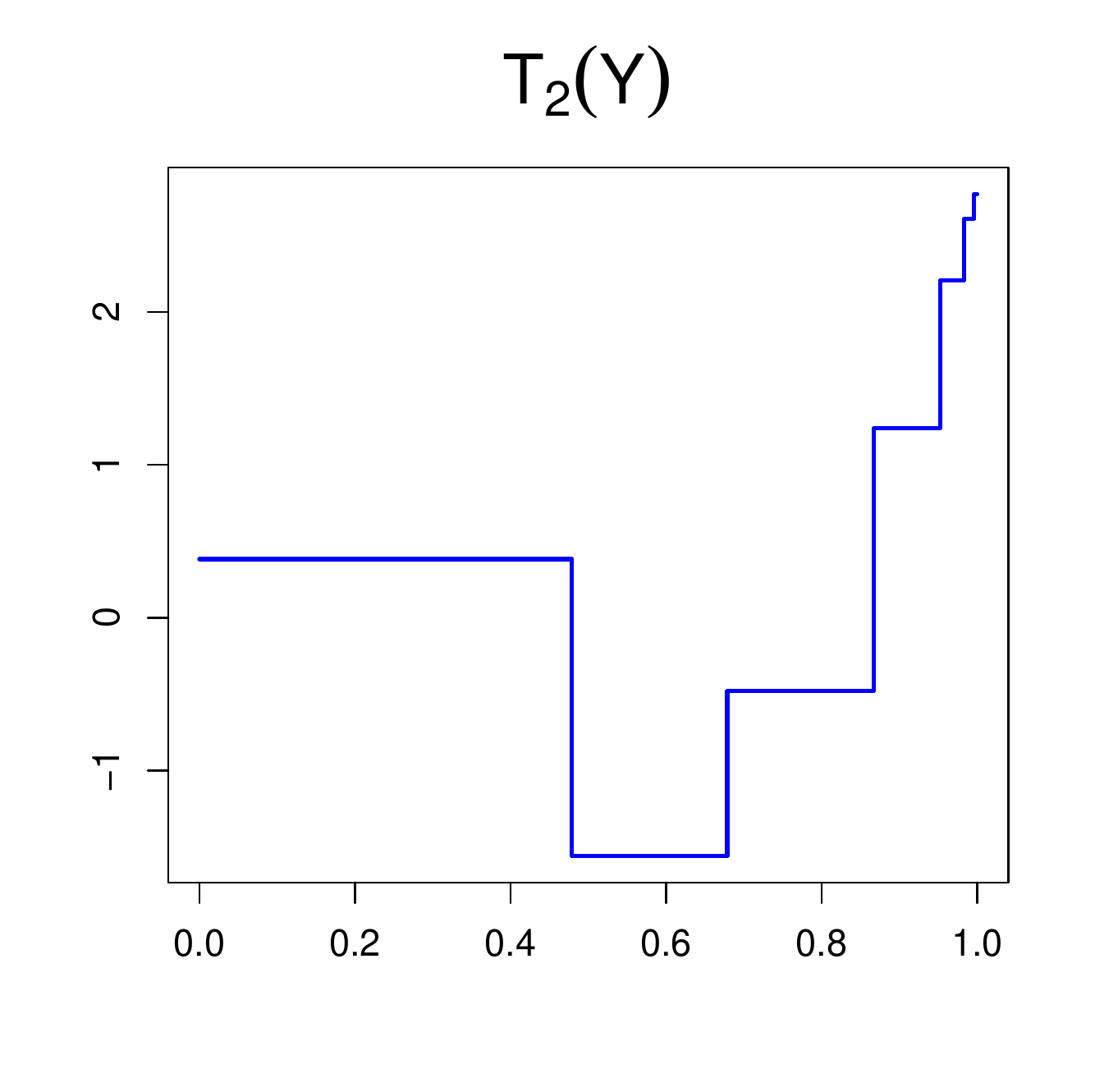}~\,
   \includegraphics[width=.23\textwidth,trim=1cm 1cm 1cm 1cm]{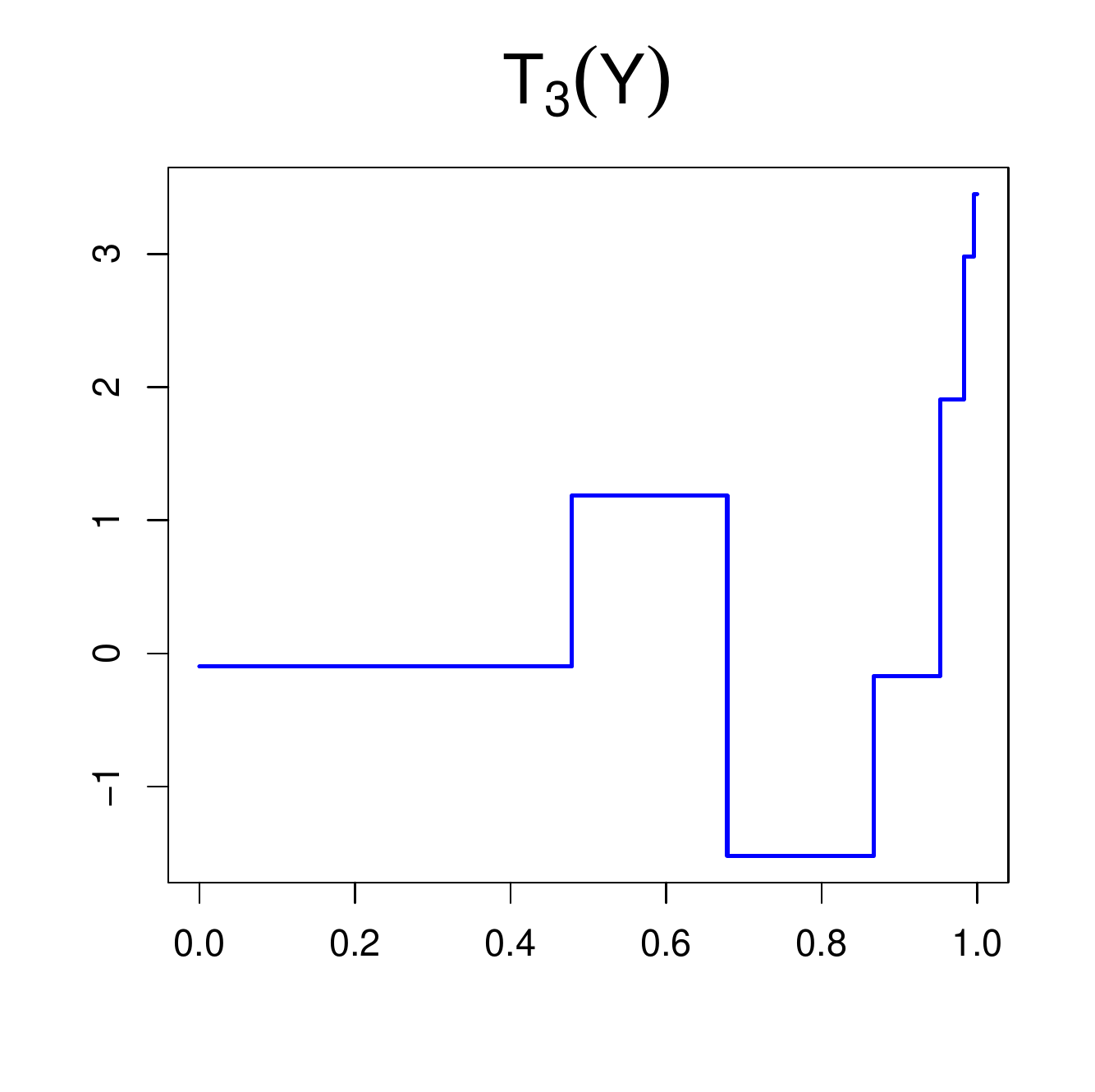}~\,
    \includegraphics[width=.23\textwidth,trim=1cm 1cm 1cm 1cm]{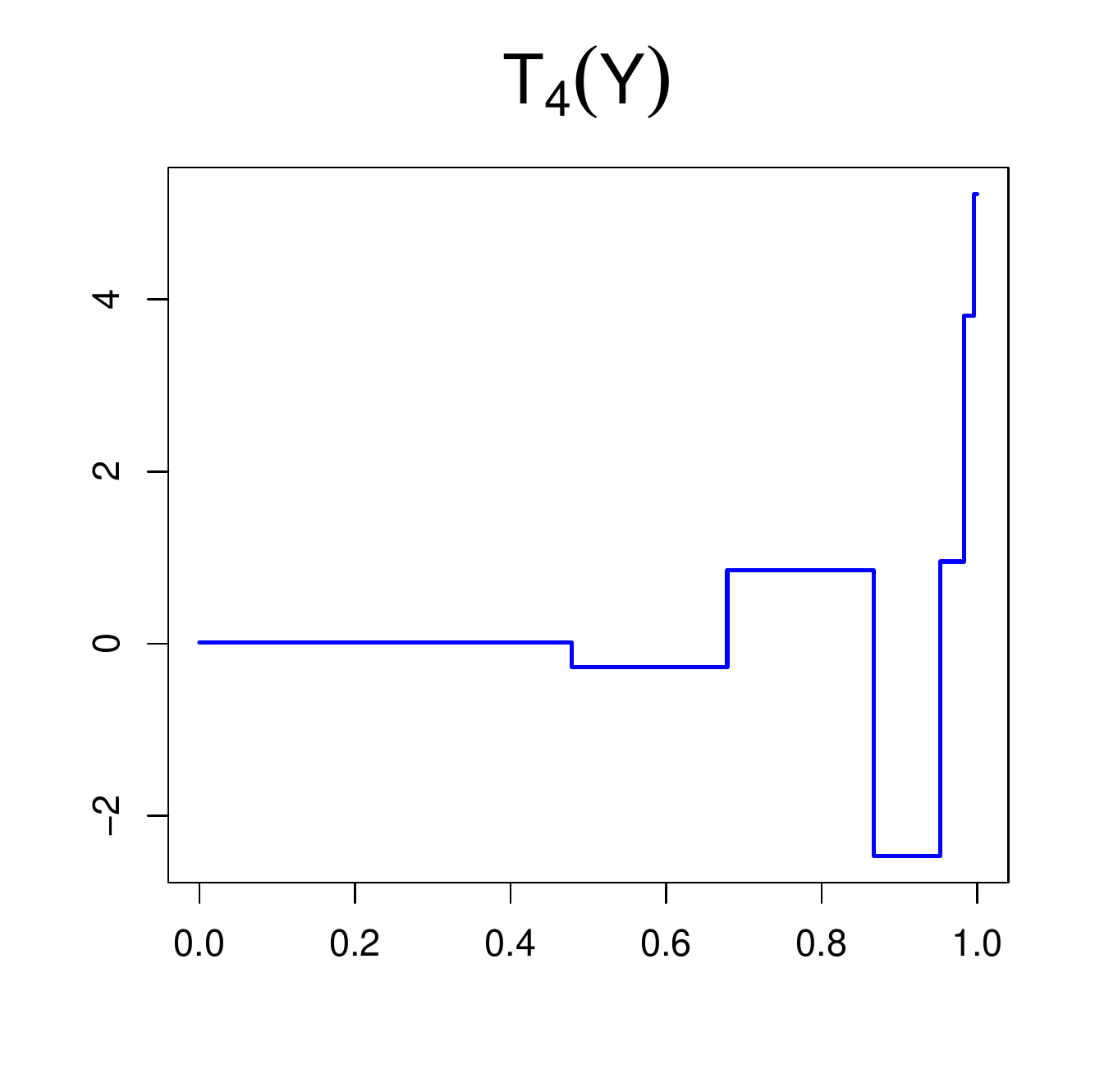}
\vskip.3em
\caption{(color online) The shapes of the first four  unit \texttt{eLP}-basis functions for wage (variable type continuous; first row in red), and (b) number of dependents (variable type discrete count; second row in blue) of Wooldridge's 1976 U.S. population survey data.}
\label{fig:sfs}
\end{figure}

{\bf Step 1:  LP-Polynomials of Mid-Ranks}. As is evident from Eq. \eqref{eq:pob}, to model copula, we need to analyze \textit{discrete} sample distributions $\wtF(x;X)$ and $\wtF(y;Y)$. Construct the LP-polynomial basis $\{T_j(X;F_X)\}_{j\ge 1}$ for the Hilbert space $\sL^2(F)$ by applying Gram-Schmidt orthonormalization (see Supplementary Appendix A for more details) on the set of functions of the power of $T_1(X;F_X)$: 
\beq \label{eq:LP1st}
T_1(x;F_X)~=~\dfrac{\sqrt{12}\big\{\Fm(x;F_X) - 1/2\big\}}{\sqrt{1-\sum_x p^3(x;F_X)}},\eeq
LP-bases obey the following orthonormality conditions with respect to the measure $F$:
\[ \int T_j(x;F_X) \dd F(x;X) =0,\,~~\text{and} \,\,\int T_j(x;F_X) T_k(x;F_X) \dd F(x;X) =\delta_{jk}.\]
For data analysis, construct the empirical LP basis (in short \texttt{eLP} basis) $\{T_j(x;\wtF_X)\}_{j=1,2\ldots,m}$, where $m$ is strictly less than the number of unique values in the sample $\{X_1,\ldots, X_n\}$. Note that our custom-constructed basis functions are orthonormal polynomials of mid-rank transform--for more details see \cite{Deep18USA,Deep17ksample}, and \cite{Deep14LP}. LP-orthonormal system plays a fundamental role in constructing expansions of our \texttt{mixed} generalized copula distributions \eqref{eq:copG}.
\begin{rem} \label{rem:npPoly}
It is important to distinguish between parametric and nonparametric (or data-adaptive) orthonormal polynomials in order to appreciate the usefulness of \texttt{eLP}-basis functions.  Traditional approaches construct separate orthonormal polynomials for standard parametric distributions like normal, exponential, Poisson, binomial, geometric, etc. on a case-by-case basis, each time solving the heavy-duty Emerson recurrence relation \citep{emerson1968}. For details, see \citet[Appendix 1]{Rayner89} and \cite{griffiths2014}. There are two practical disadvantages of this `parametric' approach:  (i) it quickly becomes fruitless (analytically laborious and computationally complex) for non-standard distributions; (ii) for real data analysis, this strategy completely breaks down, as we rarely know the underlying distribution. What we need is a `nonparametric' mechanism that can provide an \textit{automatic and universal} construction of tailor-made orthonormal polynomials for \emph{arbitrary} $X$. Applied data scientists would derive comfort from the fact that our LP-system of nonparametric polynomials serves exactly that purpose. 
\end{rem}
{\bf Step 2:  Unit LP-Basis}. Define the unit interval LP-polynomial basis functions 
\beq \label{eq:ulp}
S_j(u;F_X) = T_j\big[Q(u;X); F_X \big], ~~0<u<1
\eeq 
where $Q(u; X) =\inf\{x: F(x;X) \ge u\}$ refers to the quantile function of a random variable $X$. The eLP-unit-bases are denoted simply as $S_j(u;X)=S_j(u;\wtF_X)$. Fig. \ref{fig:sfs} displays $S_j(u;X), 0<u<1$ and $S_k(v;Y), 0<v<1$ for $j,k=1,\ldots,4$ for the Wooldridge's data. Notice the changing shape of LP-basis functions (the rows of Fig. \ref{fig:sfs})--a key characteristic property of nonparametric polynomials, as mention in Remark \ref{rem:npPoly}.
\vskip.6em

\vskip.2em
{\bf Step 3:  LP-Comeans and LPINFOR}. The next vital ingredients of our copula density estimation are LP-comeans defined as
\beq \label{LPcodef}
\LP[j,k;X,Y]=\Ex[T_j(X;F_X) T_k(Y;F_Y)],~~\text{for}~j,k>0
\eeq 
where expectation is taken with respect to the joint distribution of $(X,Y)$. The LP-comeans are the \textit{nonparametrically derived} ``parameters'' of our copula model, which can be interpreted as a dependence measure. For example, many traditional nonparametric statistics (Spearman rank correlation, Wilcoxon two-sample rank sum statistics, Pearson's phi coefficient) are equivalent to $\LP(1,1;X,Y)$; for more details see Section \ref{sec:ber} and \cite{Deep18USA}. Given bivariate data, compute and display matrix of empirical LP-comeans  (see Table \ref{WAIS:tab}) $\LP(j, k; X, Y)$  for $j, k = 1,\ldots,m$.

\vskip.5em

{\bf Step 4: LP Representation of Copula Density}.
LP orthogonal series representation of generalized copula density function for \textit{arbitrary} (X,Y) is given by
\beq \label{eq:LPCop}
\cop(u,v;X,Y)-1\,=\,\sum_{j,k>0} \LP[j,k;X,Y] S_j(u;X) S_k(v;Y),~~0<u,v<1,
\eeq
or, equivalently, LP-comeans are orthogonal coefficients of $\sL^2$ representations (estimators) of square integrable copula density
\beq \label{eq:LPmat} \int_{[0,1]^2}  \dd u \dd v\,\cop(u,v;X,Y) S_j(u;X) S_k(v;Y)\,=\,\LP[j,k;X,Y].
\vspace{-.1em}
\eeq
A proof of copula LP representation is provided by representations of conditional copula density and conditional expectations:

\beas d\big(v;Y, Y|X=Q(u;X)\big)&=&\sum_k S_k(v;Y) \Ex\big[T_k(Y;F_Y)\mid X=Q(u;X)\big];  \\
\Ex\big[T_k(Y;F_Y)\mid X=Q(u;X)\big] &=& \sum_j S_j(u;X) \Ex\big[T_j(X;F_X) T_k(Y;F_Y)\big].\eeas
LP-representation theory provides a `smooth' copula density estimate for \texttt{mixed} (X,Y) where the LP-comeans and custom-built orthogonal polynomials play the vital role. 

\vskip.5em
{\bf Step 5: Estimation and Denoising}. To estimate the LP-comeans, first note that Eq. \eqref{eq:LPmat} can be rewritten as follows after substituting $u=F_X(x)$ and $v=F_Y(y)$:
\beq \label{eq:LPalt} \LP[j,k;X,Y] = \int_{x,y} T_j(x;F_X) T_k(y;F_Y) \dd F_{X,Y}(x,y),\eeq
since $T_j(Q(u;X);F_X)=S_j(u;X)$ and $T_k(Q(v;X);F_Y)=S_j(v;Y)$ by construction. The expression \eqref{eq:LPalt}  immediately leads to the following empirical estimate of the LP-comeans:
\beq \label{eq:LPem} \tLP[j,k;X,Y] \,=\, \Ex\big[T_j(X;\wtF_X)T_k(Y;\wtF_Y); \wtF_{X,Y}\big] \,=\,\dfrac{1}{n}\sum_{i=1}^n T_j(x_i;\wtF_X) T_k(y_i;\wtF_Y).\eeq
Using the theory of linear rank statistic \citep{ruymgaart1974,D13a}, one can easily show that the sampling distribution (under the initial hypothesis of independence) of the empirical $\tLP(j,k; X,Y)$ are i.i.d $\cN(0,1/n)$. Identify the significantly non-zero (``dominant'' components) indices $j,k$ of $\tLP(j,k; X,Y)$ by using the Schwarz model selection criterion, applied to the LP-comeans arranged in decreasing magnitude:
\[\mbox{BIC}(m) = \text{Sum of squares of first $m$ comeans} - \log(n) m/n.~~~\]
Choose $m$ to maximize $\mbox{BIC}(m)$ for identifying the important LP-comeans. For other variants of penalty see \cite{kallenberg2008,Deep17LPMode}, and references therein.
\vskip.5em

{\bf Step 6: LP-Spectral Expansion of Copula Density}. 
Here we provide an alternative way of expressing (Karhunen-Lo{\'e}ve-type canonical representation) the fundamental LP-Fourier series expansion result \eqref{eq:LPCop}.

Our canonical expansion is based on the singular value decomposition (SVD) of the LP-comean kernel  \eqref{LPcodef} $\LP=U\Lambda V^T$, where $u_{ij}$ and $v_{ij}$ are the elements of the singular vectors with singular values $\la_0=1 > \la_1 \ge \la_2 \ge \cdots \ge 0$. For a  square integrable copula density, we have the following orthogonal expansion result:
\beq \label{eq:lpspec}
\cop(u,v;X,Y)-1~=~ \sum\nolimits_{k>0} \la_k \,\phi_k(u;X)\,\psi_k(v;Y),~~0<u,v<1,
\eeq
\vspace{-.4em}
where the spectral basis functions are linear combinations of LP-polynomials: $\phi_k(u;X)\,=\,\sum_{j}u_{jk}S_{j}(u;X)$, and  $\psi_k(u;Y)\,=\,\sum_{l}v_{lk}S_{l}(v;Y)$. We call these spectral-bases as copula-principal components--a potential tool for non-linear dimension reduction for mixed data.

\begin{rem}[Trivariate Copula]
The LP-nonparametric theory of copula allows extension to higher-dimension. For example, a  trivriate Bernoulli copula\footnote[2]{Modeling multivariate binary data is an important problem in economics and health  care.} admits the following representation for $0<u_1,u_2,u_3<1$:
\vskip.5em
{\small
$\cop(u_1,u_2,u_3;X_1,X_2,X_3)\,=\,1\,+\,\LP[1,1,0]S_1(u_1;X_1)S_1(u_2;X_2) \,+\, \LP[1,0,1]S_1(u_1;X_1)S_1(u_3;X_3)$\\[.25em]
$~~~~~~~~~~~~+ \LP[0,1,1]S_1(u_2;X_2)S_1(u_3;X_3)+\LP[1,1,1]S_1(u_1;X_1)S_1(u_2;X_2)S_1(u_3;X_3),
$
}
\vskip.5em
where the higher-order LP-comeans are defined as follows:
\vskip.5em
~~~~~~~~~~~~~~$\LP[j,k,l;X_1,X_2,X_3]~=~\Ex[T_j(X_1;F_{X_1}) T_k(X_2;F_{X_2}) T_l(X_3;F_{X_3})].$
\vskip.5em
As an anonymous reviewer pointed out, this multivariate LP-expansion result could be of ``great importance,'' since it is validity does not require assumptions like conditional independencies or constant conditional copulas. Some strategies on higher-dimensional generalizations are discussed in the next section.
\end{rem}
\section{Nonlinear Mixed Dependence Measure} 
How can we develop a copula-based nonparametric dependence measure that is: (i) capable of detecting complex nonlinear relationships between $X$ and $Y$, (ii) valid for mixed pairs of random variables, (iii) computationally fast enough to handle large datasets, and finally, (iv) able to provide insights into the `nature' of the non-linear pattern that is present in the data. As a solution to this problem, we introduce a new \textit{class} of dependence measure based on LP-comean matrix.

\begin{defn}
Define LP-copula based nonparametric dependence measure LPINFOR
\beq \label{eq:lpinfor} 
\LPINFOR(X,Y)\,=\iint_{[0,1]^2} \big[\cop(u,v;X,Y) -1\big]^2 \dd u \dd \,v= \sum_{j,k>0} \Big|\LP[j,k;X,Y]\Big|^2.
\eeq 
A few notable properties: (i) $X$ and $Y$ are independent \textit{if and only if} $\LPINFOR(X,Y)= 0$. (ii) It is invariant under monotone transformations (e.g., a logarithm/exponential) of the variables. (iii) Our LPINFOR statistic measures distance between true joint distribution and the independence model. Estimate LPINFOR by influential product LP-basis functions determined by BIC (or AIC).
\end{defn}

\begin{rem}[Information Measure]
The name LPINFOR arises from the observation that it can be interpreted as an INFORmation-theoretic measure belonging to the family of Csiszar's f-divergence measures \citep{csiszar1975}.
\end{rem}

\begin{rem}[Higher-dimensional Tree-Copula]
Our methodology provides rapid constriction of a multivariate dependence tree. It consists of two steps:  

\begin{itemize}[itemsep=3pt,topsep=1.4pt,leftmargin=10pt]
\item[1.] Infer the maximum spanning tree (MST) by using the LPINFOR statistic as the edge-weight (indicating the degree of dependence between two variables) where $\mathcal{E}$ denotes the $d-1$ edge sets.
\item[2.] Approximate the multivariate copula density by:
\beq  \cop(u_1,\ldots, u_d;X_1,\ldots,X_d)\, = \prod_{(i,j) \in \mathcal{E}} \cop(u_i,u_j;X_i,X_j), ~~~~\eeq
where each $d-1$ local bivariate copulas $\cop(u_i,u_j;X_i,X_j)$ are estimated using the LP-method to ensure that the whole process can run automatically for mixed variables. 
\end{itemize}
\end{rem}
\vspace{-.4em}

\begin{figure}[t]
  \centering
  \includegraphics[width=.46\linewidth,keepaspectratio,trim=.5cm .5cm .5cm .5cm]{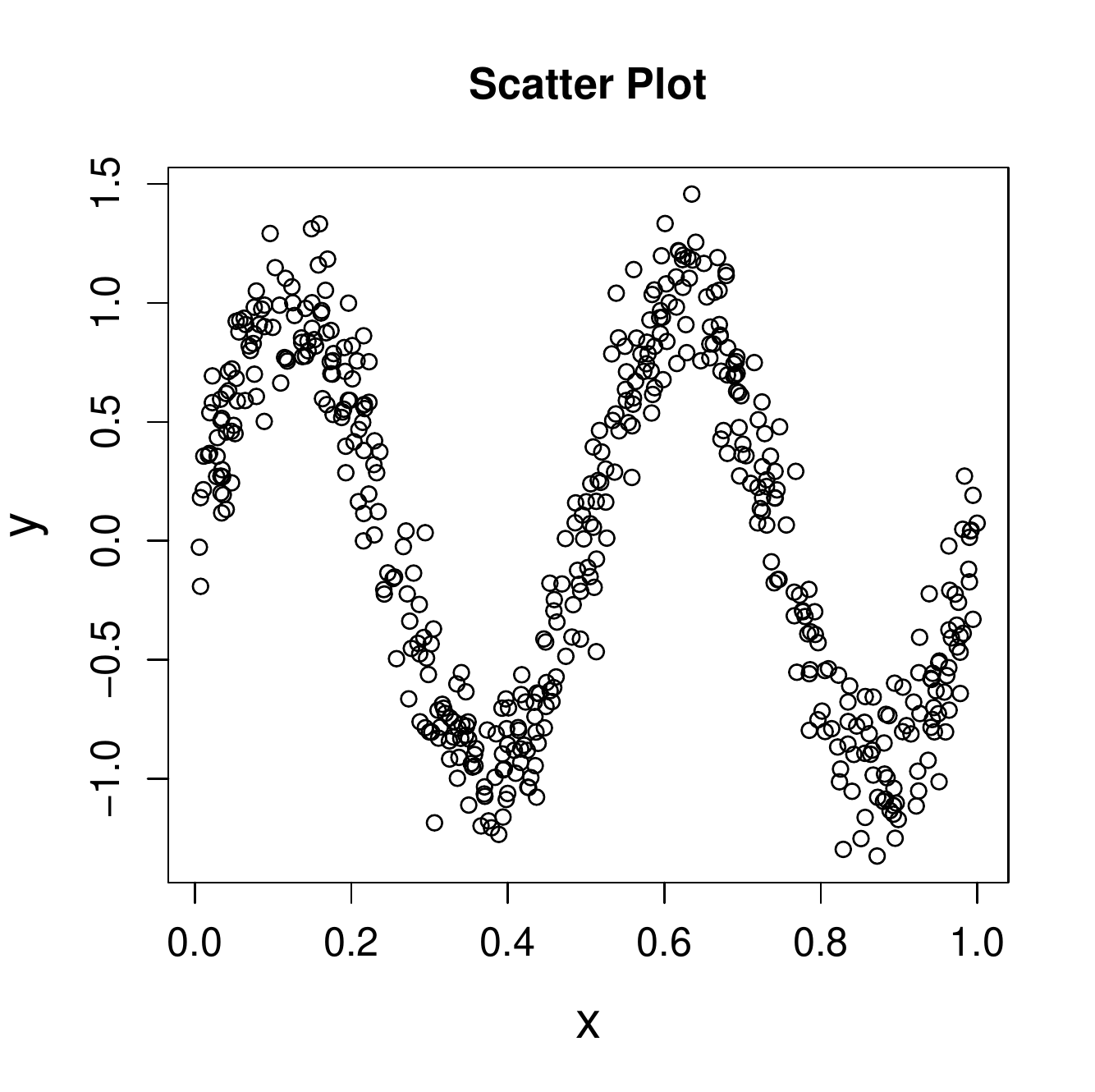}~~~~
\includegraphics[width=.46\linewidth,keepaspectratio,trim=.5cm .5cm .5cm .5cm]{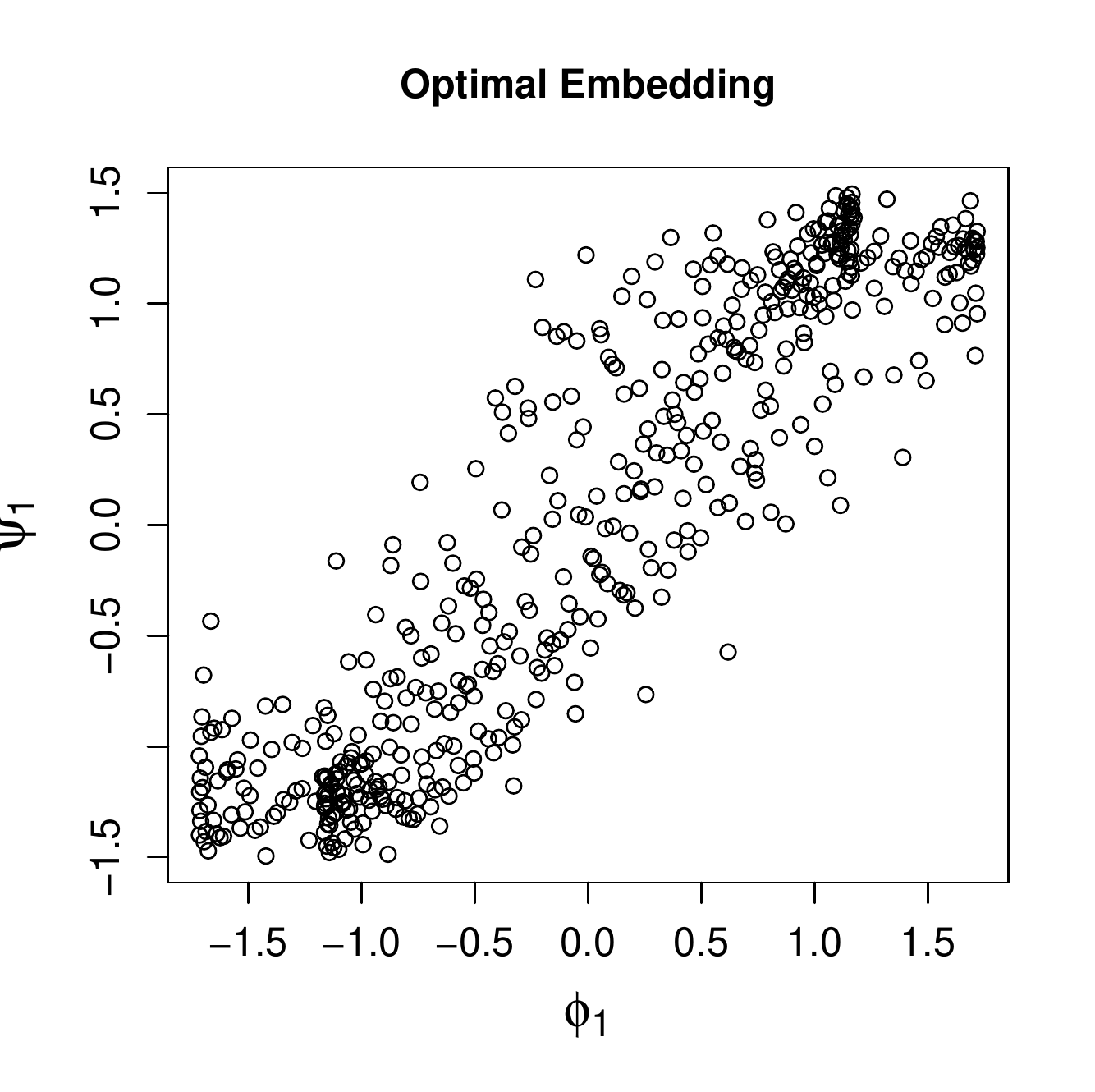} \\[1.25em]
  \includegraphics[width=.46\linewidth,keepaspectratio,trim=.5cm .5cm .5cm .5cm]{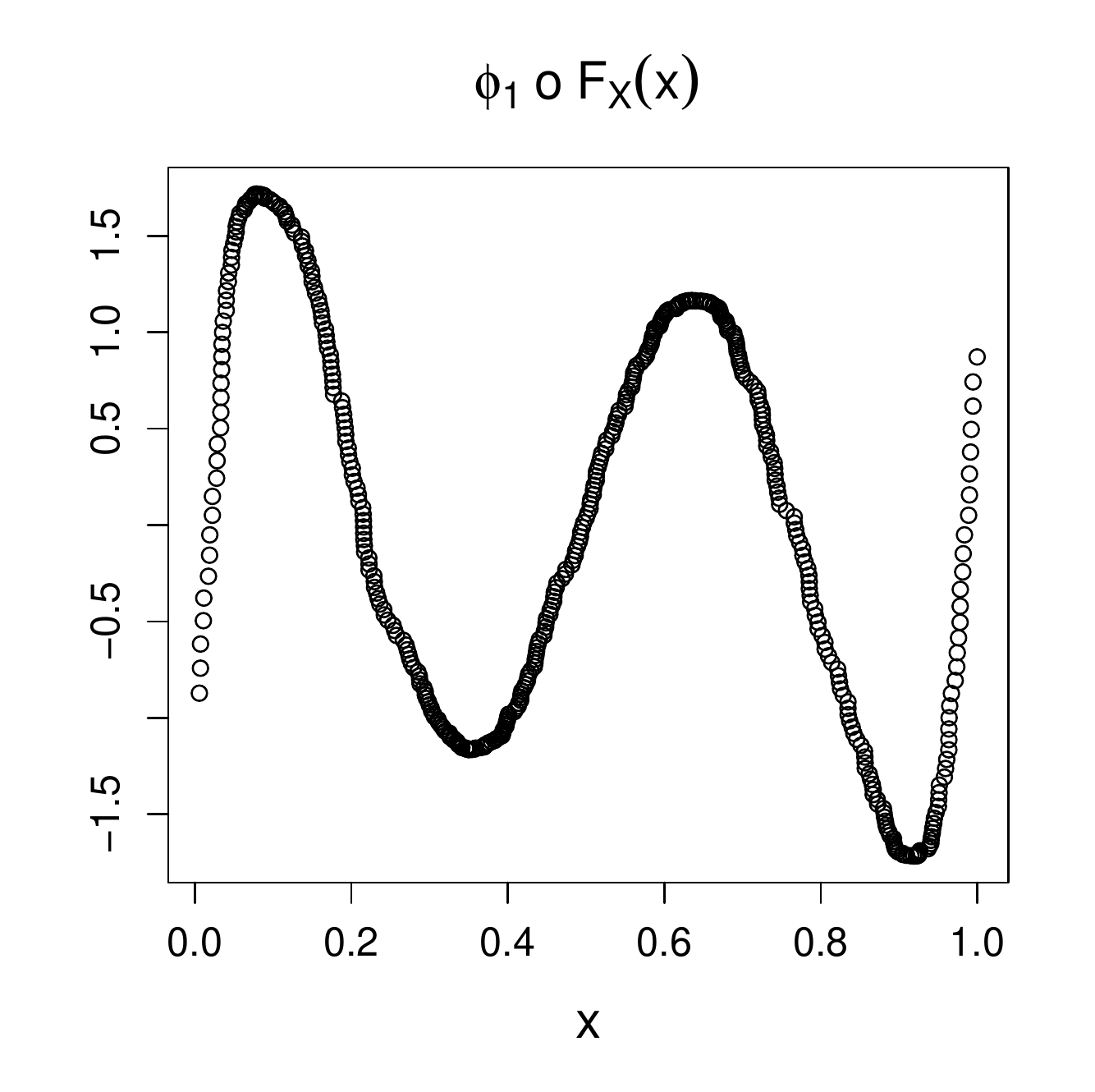} ~~~~
  \includegraphics[width=.46\linewidth,keepaspectratio,trim=.5cm .5cm .5cm .5cm]{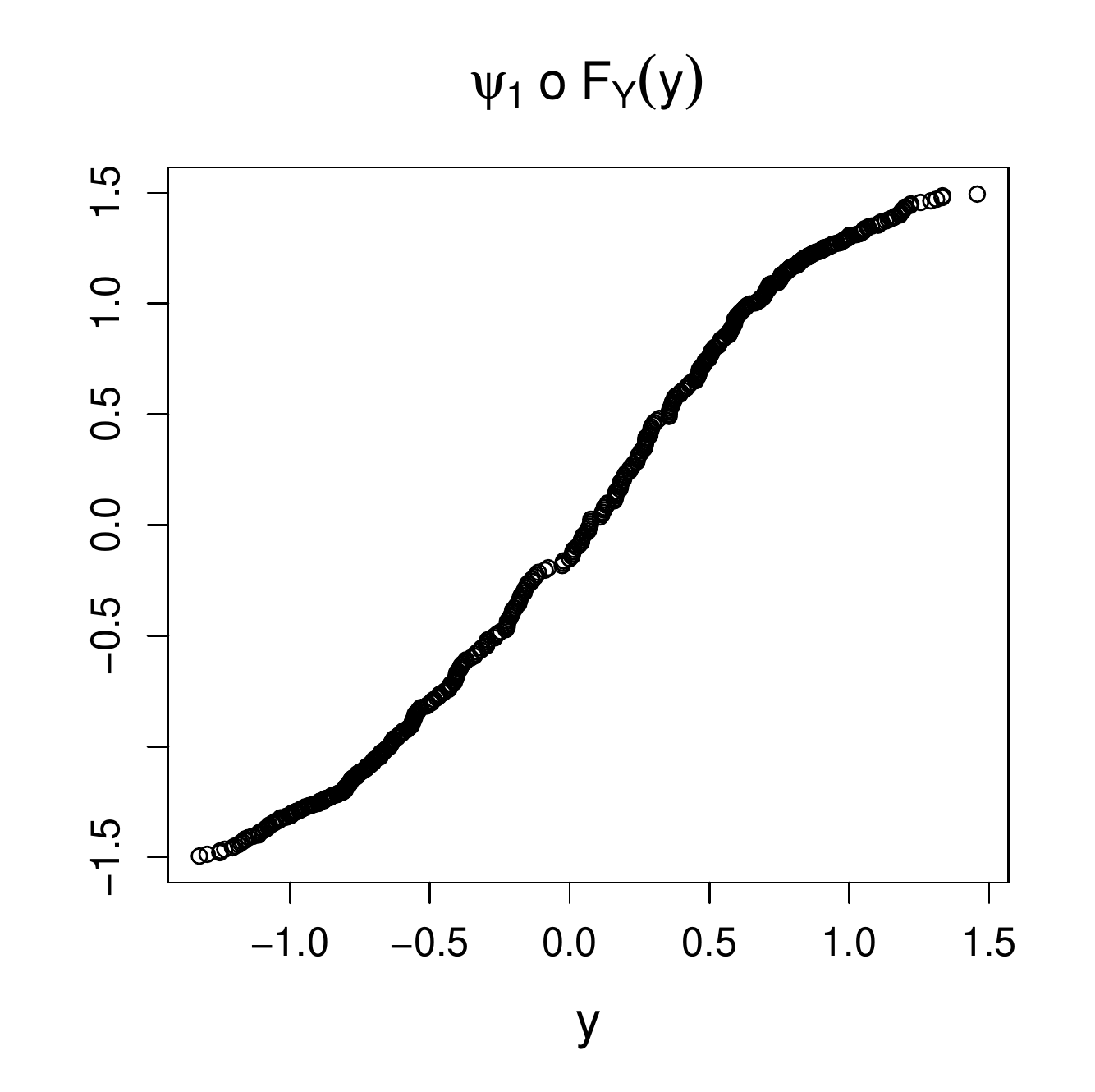} 
\caption{Example 5. (top left) The sinusoidal scatter plot ; (top right) The optimal transformation-based embedding; (bottom panel) The estimated $\phi_1 \circ \wtF_X(x)$ and $\psi_1 \circ \wtF_Y(y)$ that correctly capture the shape of the functional relationship.} 
\label{fig:LPopt}
\vspace{-.5em}
\end{figure}

{\bf Example 5}. \textit{Sinusoidal pattern}: We consider the model $y = {\rm sine}(4\pi x) + \epsilon$ with error $\epsilon\sim \cN(0,\si^2)$. Fig \ref{fig:LPopt} shows the relationship based on a sample of size $n=500$ generated from the model with $\sigma=0.40$. The associate empirical LP-comean matrix is given by

{\small\[\widehat{\LP}\,=\,\begin{bmatrix}
 0.95^*  &0.00 &-0.04  &0.02  &0.00 &-0.01\\
 0.00  &0.81^*  &0.03 &-0.08  &0.02  &0.01\\
 -0.04  &0.04  &0.64^*  &0.04 &-0.11^*  &0.00\\
 0.03 &-0.09  &0.07  &0.48^*  &0.02 &-0.1\\
 0.00 & 0.05 &-0.13^*  &0.07  &0.33^*  &0.01\\
 0.00  &0.00  &0.06 &-0.14^*  &0.06  & 0.26^*
\end{bmatrix}
\]}\vskip.15em
This yields the LPINFOR statistic value $2.51$ with p-value essentially zero, as $n \LPINFOR$ follows $\chi^2_{6 \times 6}$ under independence. However, an applied researcher might want to go beyond confirmatory test. In particular, the question of  `how' $X$ and $Y$ are dependent (the nature of nonlinear coupling) could help domain scientists to generate refined hypothesis to investigate. To bring this exploratory side into our analysis, we introduce the concept of LP-maximal correlation and optimal transformations.

\begin{defn}
Define LP-maximal correlation between $X$ and $Y$ as
\beq \label{eq:LPmax}
{\rm LPMax}(X,Y)~=~\max_{\xi,\eta} {\rm Corr} \big( \xi(X),\, \eta(Y)\big)~=~{\rm Corr}\big(  \phi_1 \circ F_X(X),\,   \psi_1 \circ F_Y(Y) \big),~~~
\eeq
where `$\circ$' denotes the usual composition of functions; $\phi_1$ and $\psi_1$ were defined in \eqref{eq:lpspec}.
\end{defn}

\begin{rem}
${\rm LPMax}(X,Y)$ not only extends \cite{breiman85} to the mixed $(X,Y)$ case, but also provides a more computationally friendlier scheme.
\end{rem}
Fig. \ref{fig:LPopt} shows the estimated optimal transformations and their embedding with the empirical ${\rm LPMax}(X,Y)=0.912$. The remarkable thing about our algorithm is how accurately it recovers the optimal transformations: the sinusoidal function of $X$ and almost linear function for $Y$, and above all the simplicity of the computation (requiring only the SVD of the LP-dependence matrix; see Step 6 of Sec 4.3). 

\begin{rem}
Our LP-theoretic approach provides copula-based unified algorithms to measure as well as explore complex correlations, without any significant computational cost. Another approach of inferring the `shape' of dependency will be discussed next.
\end{rem}
\vspace{-1em}
\section{Exploratory Modeling: The `Shape' of Dependence}
The LP-copula approach provides a model and explanation both. The LP-comeans provide insights into `how' $X$ and $Y$ are dependent. In order to better understand the exploratory side of our approach, consider the Wechsler Intelligence Scale (WAIS) data in Table 1. The question of particular interest to us: Whether age (discrete variable) and IQ (continuous variable) are dependent? If so, what can we say about the nature of their relationship? 

\begin{figure}[t]
\vskip1em
\begin{subfigure}{.46\textwidth}
  \centering
  \includegraphics[width=1.27\linewidth,keepaspectratio,trim=1.85cm 1.25cm .25cm 1cm]{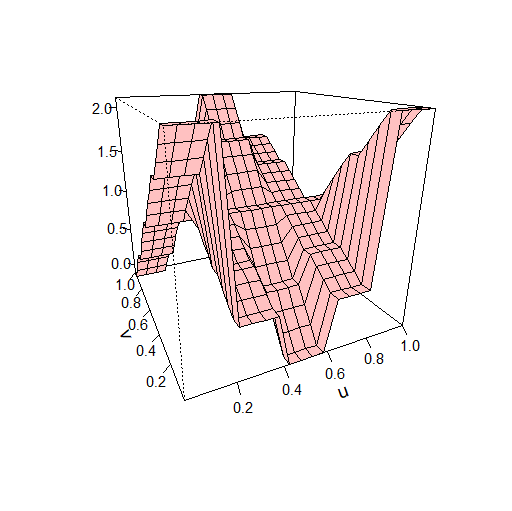}
\end{subfigure}\hspace{10mm}%
\begin{subfigure}{.45\textwidth}
\includegraphics[width=\linewidth,keepaspectratio, trim=.1cm 1cm 2.5cm 1.5cm]{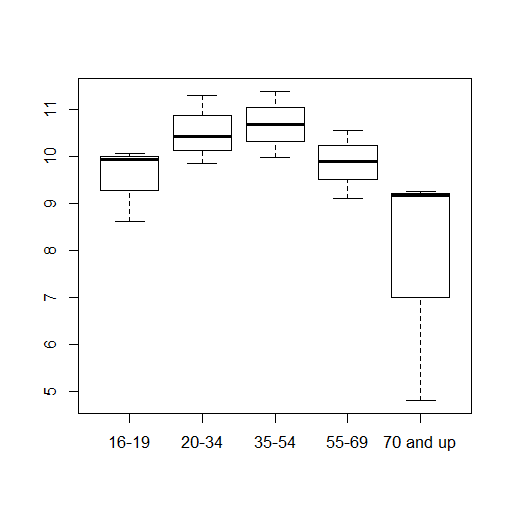}
\end{subfigure}~~~~~~

\caption{WAIS data: The estimated LP-nonparametric mixed copula estimate and the boxplot of Age and IQ.}
\label{fig:WAIS}
\end{figure}

\begin{table}[ht]
\vskip.5em
\begin{minipage}[b]{.44\textwidth}
\centering
\def\arraystretch{1.15}
\setlength{\tabcolsep}{7pt}
\centering
\begin{tabular}{c c c c c}
\hline
16-19 & 20-34 & 35-54 & 55-69 & $\geq$70\\
\hline
8.62  &9.85 & 9.98 & 9.12& 4.80\\
9.94 &10.43 &10.69  &9.89& 9.18\\
10.06& 11.31 &11.40& 10.57 &9.27\\
\hline
\end{tabular}
\end{minipage}~~~~~
\begin{minipage}[b]{.4\textwidth}
\centering
\[\widehat{\LP}\,=\,\begin{bmatrix}
 -0.32  &0.17  &0.17 &-0.11\\[.125em]
 -{\bf 0.618}^*& -0.03 &-0.10  &0.07\\[.125em]
 0.09  &0.14  &0.08  &0.04\\[.125em]
 0.16  &0.21  &0.04  &0.29
\end{bmatrix}
\]
\end{minipage}
\vskip.65em
\caption{The table on the left displays the IQ score of $n = 15$ adults taking the Wechsler Intelligence Test \citep[Ch. 6]{hollander2013book} and, at right, the estimated LP-dependence matrix.} \label{WAIS:tab}
\end{table}
The LP-comean matrix, displayed in Table \ref{WAIS:tab}, immediately implies the following: 
\begin{itemize}[itemsep=3pt,topsep=1.4pt,leftmargin=10pt]
\vskip1em
\item There exists a significant (non-linear) dependence between Age and IQ score (LPINFOR-based p-value being $0.00835$).
\item The only significant LP-component is $\LP[2,1;\text{Age}, \text{IQ}]=-0.618$, which indicates interaction between linear-IQ and quadratic-age effect.
\item In addition, the negative sign of $\LP[2,1;\text{Age}, \text{IQ}]$ suggests there is an umbrella-like trend of IQ as a function of age. This statistically confirms the common belief that the ability to comprehend ideas and learn is an increasing function of age up to a certain point, and then it declines with increasing age. This is also verified from the boxplot in Fig. \ref{fig:WAIS}. We feel this exploratory side of our method could be valuable for applied users to better interpret the pattern in the data.
\item The final estimated (\textit{mixed}: IQ score is continuous; Age groups is a discrete variable) LP-copula density is shown in the left panel of Fig. \ref{fig:WAIS}.
\end{itemize}

\section{Bernoulli Copula: Paradox and Its Resolution} \label{sec:ber}
\begin{quote}
{\small ``\emph{Surely discrete marginal dfs $F_1, \ldots, F_d$ must cause problems. Believe me, they do! be prepared that everything that can go wrong, will go wrong}.''}---\cite{embrechts2009}.
\end{quote}
\vspace{-.2em}
Developing a nonparametric copula-based dependence measure for discrete data (i.e., where the probability of a tie is positive) is known to be a challenging problem. However, an even more ambitious target is to design a \textit{truly automated} correlation-learning algorithm for mixed variables that can adapt itself intelligently \textit{without} having the data-type information from the user.

\vskip.65em
The problem comes from the fact that, while concordance measures like Kendall's $\tau$  or Spearman's $\rho$ are margin-free for continuous random variables, the same does not hold for discrete cases. As a consequence, the range of copula-based correlation measures for discrete $(X,Y)$ become \emph{marginal-dependent}, and do not attain the bound $\pm 1$. This creates a significant complication, since we can not \emph{compare the `strength' of dependence} by solely comparing it's magnitude. The following example, taken from \citet[p. 492]{genest2007} illustrates this non-intuitive unpleasant phenomena. 
\vskip.5em

{\bf Example 6}. \textit{The Bernoulli table paradox}:
Consider $(X,Y)$ Bernoulli random variables with marginal and joint distribution $\Pr(X=0)=\Pr(Y=0)=\Pr(X=0,Y=0)=p \in (0,1)$, which implies $Y=X$ almost surely, still the traditional ${\rm Spearman}(X,Y)=p(1-p)<1$.  Similarly, for the case $\Pr(X=0)=\Pr(Y=1)=\Pr(X=0,Y=1)=p \in (0,1)$, despite $Y=1-X$ almost surely, the traditional ${\rm Spearman}(X,Y)=-p(1-p)>-1$ yields counter-intuitive  answer. This shows the surprising phenomenon that discrete $X$ and $Y$ with perfect monotone functional dependence does not guarantee  $|{\rm Spearman}(X,Y)|= 1$.

\vskip.5em
{\bf Generalized Spearman Correlation.} 
We now show the $\LP[1, 1; X, Y ]$ (linear-rank) statistic not only resolves these inconsistencies but also provides one \textit{single computing formula} for Spearman correlation that is valid for mixed (discrete or continuous) marginals without any additional adjustments.

\vskip.6em
\textit{Step 1.} For binary $X$ with $\Pr(X=0)=p$: we have $\Fm(0;F_X)=p/2,$ $\Fm(1;F_X)=(1+p)/2,$ and the correction factor $\sqrt{1-\sum_{x=0}^1 p^3(x;F_X)}=\sqrt{3p(1-p)}$.
\vskip.4em
\textit{Step 2.} Thus we have the mid-distribution-based LP-polynomial: $T_1(0;F_X)=-\sqrt{\frac{1-p}{p}}$, and $T_1(1;F_X)=\sqrt{\frac{p}{1-p}}$. Remember that as $X$ is binary (takes two distinct values), we have one LP-basis function.
\vskip.4em
\textit{Step 3.} The LP-basis for $Y$ is $T_1(0;F_Y)=-\sqrt{\frac{1-p}{p}}$, and $T_1(1;F_Y)=\sqrt{\frac{p}{1-p}}$, since $Y$ has the same marginals as $X$ with $\Pr(Y=0)=p$.
\vskip.4em
\textit{Step 4.} Now we are in a position to compute the $\LP[1,1;X,Y]$:
\vspace{-.5em}
\begin{align*}
\LP[1,1;X,Y] ~&=~ \Ex[T_1(X;F_X) T_1(Y;F_Y)]\\ 
~&=~p T_1(0;F_X) T_1(0;F_Y) + (1-p)T_1(1;F_X) T_1(1;F_Y)\\
~&~= 1.
\vspace{-.75em}
\end{align*}
A similar calculation goes through for the second case, where we have $\Pr(X=0)=\Pr(Y=1)=\Pr(X=0,Y=1)=p \in (0,1)$, yielding $\LP[1,1;X,Y]=-1$.
\begin{rem}
In summary, generalized LP-correlations circumvent the mammoth challenges and technicalities required to build a valid copula-based dependence measure for mixed data. We end this section with a real example. 
\end{rem}

{\bf Example 7}. \textit{Hellman's Infant Data} \citep{yates1934}: Table \ref{tab:yates} shows cross-tabulation of $n=42$ infants based on whether the infant was breast-fed or bottle-fed. 
\begin{table}[h]
\vskip1em
\begin{center}
\setlength{\tabcolsep}{1.15em}
\begin{tabular}{c|cc}
& Normal teeth & Malocclusion\\
\hline\\[-.4em]
Breast-fed & 4 & 16 \\[.2em]
Bottle-fed & 1& 21\\[.2em]
\hline
\end{tabular}
\end{center}
\caption{The data table on malocclusion of the teeth in infants were obtained by M. Hellman and reported in the classic paper by Frank Yates (1934, p.230).} \label{tab:yates}
\end{table}
The scientific question of interest here is whether the type of feeding is associated with malocclusion. For this data we estimate the LP-comean:
\[\hLP[1,1;X,Y]\,=\, {\rm Cov}\big(T_1(X;\wtF_X), T_1(Y;\wtF_Y)\big)\,=\,0.238.\]
with the pvalue $1-\Phi(\sqrt{42} \times 0.238) = 0.061$, indicating an absence of any notable correlation. 

\section{Tests of Symmetry and Direction}
In the previous section, we have seen how LP-comean dependence matrix helps to uncover the functional relationship between $X$ and $Y$. Now we will venture to go one step further and ask: does LP-comean matrix inform the user of the `shapes' of copula density function? In particular, we are interested in the following question: 
\vskip.65em
\textit{Can we transform the problem of checking the symmetry of bivariate copula density} $\cop(u,v;X,Y)$ $= \cop(v,u;X,Y)$ \textit{for all} $(u,v) \in [0,1]^2$, \textit{as testing symmetry of the corresponding LP-comean matrix?}
\vskip.65em
For the purposes of illustration, consider the following two empirical LP-comean matrices, computed from the Gaussian($\rho =.8$) copula and its 
asymmetric version produced by Khoudraji's device \citep{khoudraji1996} with $(\la_1=.1, \la_2=.6)$:
\beq
\Cop^{\mbox{K}}(u,v;\la_1,\la_2)\,=\,u^{1-\la_1}v^{1-\la_2} \Cop(u^{\la_1},v^{\la_2}), ~~~\text{for $\la_i \in (0,1)$, $\la_1 \neq \la_2$}.
\eeq
\vspace{-1em}
\begin{minipage}[h]{.4\textwidth}
\centering
\[\widehat{\LP}^{{\rm Gaus}}\,=\,\begin{bmatrix}
 0.74^*  &0.01  &0.05 &-0.03\\
 -0.01  &0.58^* &-0.03  &0.11^*\\
 0.08 &-0.04  &0.36^* &-0.04\\
 -0.03  &0.10^* &-0.04  &0.22^*
\end{bmatrix}
\]
\end{minipage} \hskip4.5em
\begin{minipage}[h]{.4\textwidth}
\centering
\[\widehat{\LP}^{{\rm K}}\,=\,\begin{bmatrix}
 0.38^* &-0.02  &0.01  &0.00\\
 0.18^*  &0.23^* &-0.05  &0.01\\
 0.10  &0.11^*  &0.08 &-0.02\\
 0.01  &0.06  &0.10^*  &0.10^*
\end{bmatrix}
\]
\end{minipage}
\vskip1.65em
At a first glance, $\widehat{\LP}^{{\rm Gaus}}$ looks very close to symmetry (in a stochastic sense), whereas a clear asymmetry is visible in the $\widehat{\LP}^{{\rm K}}$ matrix. We can test the hypothesis of symmetry of the LP-comean  matrix by constructing the following statistic:
\beq \label{eq:LPsym}
\LPsym ~=~\frac{1}{2} \sum\limits_{j < k} \Big| \LP[j,k;X,Y] \, - \, \LP[k,j;X,Y] \Big|^2.
\eeq
where  $n \LPsym$ follows chisquare distribution under null. We apply this test for our toy examples, leading to p-values $0.995$ and $3.87 \times 10^{-8}$ respectively.

\vskip1em

{\bf Example 8}. \textit{Geyser data}: We are given $n=272$ observation of waiting time between eruptions and the duration of the eruption for the Old Faithful geyser in Yellowstone National Park. The left panel of Fig. \ref{fig:geyser} displays the data whose LP-comean matrix is displayed below:
\[ \hLP[\mbox{Eruptions},\mbox{Waiting}]~=~
\begin{bmatrix}
                0.780^* &-0.190^* & -0.130  &0.208^*\\
                -0.181^*  &0.290^*  &0.038 &-0.040\\
                -0.137  &0.053  &0.169^* &-0.019\\
                0.190^* &-0.096  &0.042  &0.108
                             \end{bmatrix}\]
Few notable features are clear from the LP-matrix: \begin{itemize}[itemsep=3pt,topsep=1.4pt,leftmargin=10pt]
\vskip1em
\item The significant higher-order LP-comeans indicate strong presence of nonlinear correlation. Terence Speed in IMS Bulletin 15 (March 2012 issue) asked whether the dependence between eruption duration and waiting time is linear. 
\item The matrix looks very ``close'' to symmetrical. In order to assess this, we apply the $\LPsym$ test. The resulting pvalue turns out to be $0.985$. The conclusion of symmetry is not surprising looking at the scatter plot.
\item The right panel shows the estimated LP-copula density for geyser data, which indicates the presence of tail-dependence.
\end{itemize}
\begin{figure}[t]
\vskip1em
\centering
\begin{subfigure}{.46\textwidth}
  \centering
  \includegraphics[width=\linewidth,keepaspectratio, trim=2.5cm 1cm 0cm 2.2cm]{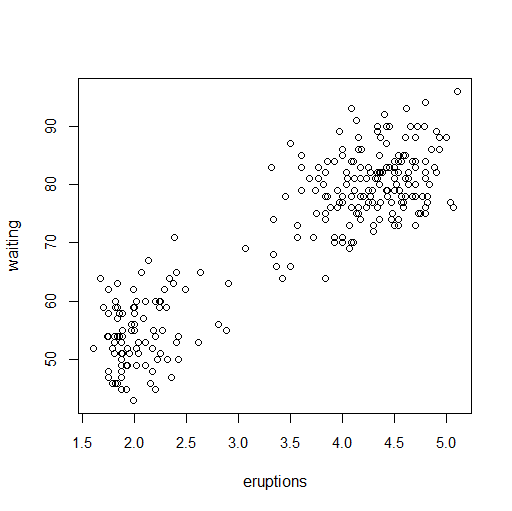}
\end{subfigure}~~~
\begin{subfigure}{.48\textwidth}
  \includegraphics[width=1.18\linewidth,keepaspectratio,trim=1.8cm 1.25cm 1.95cm .5cm]{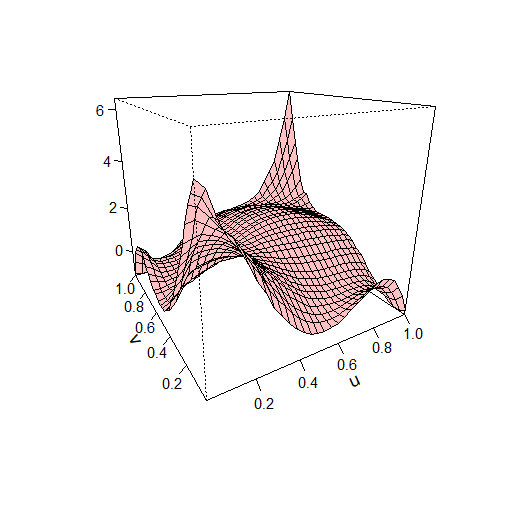}~~~~~~
\end{subfigure} 
\vspace{-1em}
\caption{Geyser data scatter plot and the LP-estimated copula density function. Symmetry is clearly visible from both.}
\label{fig:geyser}
\end{figure}

\begin{rem}
The problem of symmetry  is interesting for a few reasons: (i) this can be used to guide the suitable parametric model (e.g., Archimedean vs meta-elliptical copulas) for the data; (ii) the asymmetry might hint at a causal direction between $X$ and $Y$.
\end{rem}
\begin{rem}
LP-dependence matrix acts as both confirmatory and exploratory diagnostic tool in the sense that it contains information about the `strength' (departure from uniformity) and `shape' of the copula density.
\end{rem}
\begin{rem}
Note that, in contrast to most other schemes \citep{junker2019copula,genest12}, our approach \emph{does not require empirical estimation of copula as an intermediate problem} to understand the (a)symmetry, leading to computationally extremely efficient algorithm. This could be a huge advantage for large-scale problems. In general, LP-comean dependence matrix provides a fast and elegant way to nonparametrically diagnose different shapes of copula density. This will be explored further in the future. 
\end{rem}

\section{Simulation Study}
We perform numerical comparisons with state-of-the-art nonparametric copula density estimation methods, implemented in the R package \texttt{kdecopula} \citep{kdecopulaRpckg}:
\begin{itemize}[itemsep=1pt,topsep=1pt]
\vskip.3em
\item Probit transformation estimator with log-quadratic local likelihood estimation and nearest-neighbor bandwidths \citep{geenens2017}.
\item The classical mirror-reflection estimator proposed by \cite{gijbels1990}.
\item The penalized Bernstein polynomial-based estimator \citep{sancetta2004}.
\item  The beta kernel estimator of \cite{charpentier07}.
\end{itemize}
We restrict ourselves to continuous margins, as the competing methods are not generalizable for \texttt{mixed} data-type. Our numerical setting closely follows \citet[Sec. 5]{geenens2017}. We simulate $B=250$ independent random copula-samples $\{(U_i,V_i)\}_{i=1}^n$ of size $n=1000$ from the following copula distributions:
\begin{itemize}[itemsep=1pt,topsep=1pt]
\vskip.3em
\item Gaussian copula with parameter $\rho=0.70$;
\item Student's t-copula with 5 degrees of freedom, with  $\rho=-0.30$;
\item Frank copula with parameter $\theta=6, -2$. 
\item Plackett Copula with parameter $\theta=6, 0.10$.
\item Clayton copula with parameter $\theta=3,-0.50$.
\item Ali-Mikhail-Haq copula with parameter $\theta=0.85,-0.85$.
\item Joe copula with parameter $\theta=1.5$. 
\item Gumbel copula with parameter $\theta=1.5$. 
\end{itemize}
We assess the fit of an estimator using the mean integrated absolute error (MIAE) or $L_1$-distance criterion $\Ex\big[\int_{(0,1)^2} \big|\hcop(u,v) - \cop(u,v) \big|\dd u\dd v\big]$, which is estimated by the average over $250$ Monte Carlo replications on the grid $\{(\frac{i}{L+1}, \frac{j}{L+1})\}_{1\le i,j\le L=50}$. The results are shown in Table \ref{tbl:simu_data}. To better interpret the numbers, we have reported the MIAE of the competing methods relative to our proposed LP-method. Thus, any number greater than one indicates the superiority of the LP-copula method. 

\begin{table}[h]
\vskip.4em
\setlength{\tabcolsep}{15.5pt} 
\def\arraystretch{1.35}
\centering
\begin{tabular}{lccccc}
  \toprule
Copula family  & Probit & MR & Bernstein & Beta & LP\\
  \midrule
Gaussian (0.70)    &0.677 &1.200 &1.110 &1.221   &\cmark \\
Student's-T$_4$ (-.30)  &0.865 &1.533 &1.074 &1.187 &\cmark\\ 
Frank (6)   &1.055 &1.493 &1.811 &1.658 &\cmark\\
Frank (-2)  &1.215 &1.615 &1.423 &1.669 &\cmark\\
Plackett (6)   &1.197 &1.951 &1.860 &1.843 &\cmark\\
Plackett (0.10)  &1.576 &1.580 &1.779 &1.573 &\cmark\\
Clayton (3)   &0.688 &0.993 &1.047 &0.842 &\xmark\\
Clayton  (-0.50)  &1.282 &1.147 &0.954 &0.975 &\xmark\\
AMH (0.85)      &0.966 &1.327 &1.052 &1.240 &\cmark\\
AMH (-0.85)    &0.988  &1.012  &1.041  &1.226 &\cmark\\
Joe (1.5)       &0.875 &1.166 &0.906 &0.933&\xmark\\
Gumbel  (1.5) &0.740 &1.112 &1.095 &1.016& \cmark\\
\bottomrule
\end{tabular}  
\vskip.4em
\caption{\label{tbl:simu_data} MIAE of different nonparametric copula estimation methods relative to LP-approach. Throughout we have used $m=4$. The last column indicates whether the proposed LP-copula estimation technique is among the top two performing methods for the corresponding copula.} 
\end{table}

As we can see from Table \ref{tbl:simu_data}, more often than not, the LP-method occupies the best or second-best position. For Ali-Mikhail-Haq (AMH) copulas, all methods work equally well. For the Joe copula, the mirror-reflection kernel estimator performs quite well. For the Clayton copula family the beta-kernel estimator is very efficient. However, the probit-transformation-based kernel density method (combined with local log-quadratic approximation and K-NN-type bandwidth matrix) appears to be the most prominent competitor. All in all, considering computational efficiency, ease of implementation\footnote[2]{which, by the virtue of \eqref{eq:LPCop}, only requires computation of LP-comeans. This can be done in one line R-code: {\rm Cov}($T_X,T_Y$), where $T_X$ and $T_Y$ denote the matrix of LP-polynomials for $X$ and $Y$.}, and power of generalizability, the proposed technique has the potential to become a `default algorithm' for copula dependence modeling. 
\section{Conclusion}
The ultimate goal of copula statistical learning is to construct a simple yet flexible mathematical model with interpretable parameters that can adequately describe the essential dependence structure of the data. Taking inspiration from the recent progress on `LP-United Data Science' \citep{D13a,Deep14LP,deep16LSSD,Deep17LPMode,Deep17DMT,D12e,Deep18nature}, this article presents a modern unifying copula learning program that is valid for \textit{any} data-type. This philosophy of algorithm design is a significant milestone compared to the existing culture of building `well-tuned' retail procedures on a case-by-case basis. 
\vspace{-.45em}
\begin{quotation}
\hspace{-2em} {\small ``\textit{Efficiency for the user needs to be interpreted quite differently for the user than the tool forger.  All efficiencies between 90\% and 100\% are NEARLY the SAME for the user...The Tool-forger, on the other hand, should pay attention to another 1/2\% of efficiency.}}''{\rm --- John \citet[p. 104]{tukey1979robust}}
\end{quotation}
\vspace{-.45em}
We hope that the advances presented here will make copula-based data analysis more attractive by making it self-consistent and easy to apply for practitioners, who like to have an automated versatile tool in their statistical repository that works reasonably well for a wide range of scenarios. 
\section*{Acknowledgement}
The authors thank the editor, associate editor, and reviewers for their constructive comments and suggestions, which helped improve the paper.
\section*{Supplementary Material}
To better highlight the main points of the paper, we have relegated additional details on computation, methods, and numerical simulations to the supplement. All data used in this research are publicly available through R-software. 

\newpage
\setcounter{page}{1}
\setcounter{equation}{0}
\renewcommand{\theequation}{E.\arabic{equation}}
\renewcommand{\baselinestretch}{1.34}

\begin{center}
{\Large {\bf Supplementary Information for ``Nonparametric \vskip.15em Universal Copula Modeling''}}\\[.17in] %
Subhadeep Mukhopadhyay$^*$ and  Emanuel Parzen\\  
$^*$ To whom correspondence should be addressed; E-mail: deep@unitedstatalgo.com\\[3em]
\end{center}

This supplementary document contains three Appendices, presenting some additional methodological and numerical details.

\vskip1em
\begin{center}
{\large A. Gram-Schmidt Orthonormalization}
\end{center}
We start by explicitly defining orthonormal polynomials. As the name suggests, these are polynomials that are orthonormal to each other with respect to weighted $\sL^2$ inner product, i.e.,
\beq
\langle \xi_j, \xi_k \rangle\,=\,\int_x  \xi_j(x) \xi_k(x) \dd W(x) \,=\, \delta_{jk},~~\text{for all $j,k$.}
\eeq
Orthogonal polynomials can be obtained by applying the Gram-Schmidt orthogonalization process to the basis $\{1, \xi_1, \xi_1^2,\xi_1^3,\ldots\}$. 
Gram-Schmidt orthogonalization works as follows:
\vskip.3em
Step 1. Select $\xi_1(x)$ and the weight function $W(x)$. For constructing eLP-polynomials choose $W(x)$ to be the empirical cdf $\wtF_X$ and
\beq \xi_1(x)~\equiv~T_1(x;\wtF_X)~=~\dfrac{\sqrt{12}\big\{\Fm(x;\wtF_X) - 1/2\big\}}{\sqrt{1-\sum_x p^3(x;\wtF_X)}}.\eeq
\vskip.2em
Step 2. We then iteratively construct the next degree polynomial by removing the components in the directions of the previous ones: $T_{k+1}=\frac{T^{\diamond}_{k+1}}{\| T^{\diamond}_{k+1}\|}$, where  
\beq 
T^{\diamond}_{k+1}(x)~=~\xi_1^{k+1} ~-~ \sum_j \big \langle  \xi_1^{k+1}, T_j\big \rangle\, T_j(x;\wtF_X).
\eeq
\vspace{-.4em}
There are several functions available in \texttt{R} to perform the numerical calculation of Gram-Schmidt algorithm. Note that our custom-constructed LP-basis functions are orthonormal polynomials of mid-rank transform instead of raw $x$-values, thus provide robustness.

\newpage

\begin{center}
{\large B. Additional Simulation Results: n=500 Case}
\end{center}

\begin{table}[h]
\vskip.4em
\setlength{\tabcolsep}{15.5pt} 
\def\arraystretch{1.35}
\centering
\begin{tabular}{lccccc}
  \toprule
Copula family  & Probit & MR & Bernstein & Beta & LP\\
  \midrule
Gaussian (0.70)    &0.624  &1.217  &1.162  &1.261  &\cmark \\
Student's-T$_4$ (-.30)  &0.824  &1.329  &1.028  &1.094 &\cmark\\ 
Frank (6)   &1.051 &1.422 &1.684 &1.573&\cmark\\
Frank (-2)  &1.171 &1.477 &1.282 &1.509 &\cmark\\
Plackett (6)   &1.088 &1.762 &1.652 &1.657 &\cmark\\
Plackett (0.10)  &1.394 &1.623 &1.827 &1.620 &\cmark\\
Clayton (3)    &0.595  &0.967  &1.122  &0.930 &\xmark\\
Clayton  (-0.50)  &1.164& 1.168 &1.030 &1.051&\cmark\\
AMH (0.85)      &0.807 &1.093 &0.888 &1.029&\xmark\\
AMH (-0.85)    &1.003 &0.976 &1.035 &1.185&\cmark\\
Joe (1.5)       &0.685 &1.064 &0.767 &0.873&\xmark\\
Gumbel  (1.5) &0.869  &1.049  &1.064  &1.100& \cmark\\
\bottomrule
\end{tabular}  
\vskip.4em
\caption{ MIAE of different nonparametric copula estimation methods relative to LP-approach with $n=500$. Throughout we have used $m=4$. The last column indicates whether the proposed LP-copula estimation technique is among the top two performing methods for the corresponding copula.} 
\end{table}

\begin{center}
{\large C. Computational Time and Implementation Ease}
\end{center}

\begin{table}[h!]
\setlength{\tabcolsep}{14pt}
\def\arraystretch{1}
\vskip.5em
\centering
\begin{tabular}{ c ccccc}
\toprule
\toprule
Methods&\multicolumn{4}{c}{Size of the data sets}\\
\cmidrule(r){2-5}
&$n=500$ & $n=1000$ & $n=5000$ & $n=10,000$ \\
\midrule
Probit  &  0.3592  &0.5828  &2.4472  &4.8352\\[1ex]
MR &  0.0504  &  0.1160  & 1.1644 & 3.5976 \\[1ex]
Bernstein  & 0.2944 & 0.6560  & 4.6060 & 10.9684  \\[1ex]
Beta  & 0.4016 & 0.7676 & 3.9548 & 8.9556 \\[1ex]
LP & 0.0040 &  0.0108  &0.1916  & 0.6536 \\[1ex]
\hline
\end{tabular}
\vskip1em
\caption{Computational time: uniformly distributed, independent samples of size $n$, averaged over $50$ runs based on Intel(R) Core(TM) i5-8250U CPU @ 1.60GHz, 1800 Mhz, 4 Core(s) processor. Timings are reported in seconds.}
\label{tab:time}
\end{table}

For moderately large problems it seems: LP is almost 50x faster than Probit, 10x faster than MR, 60x faster than Bernstein, and 70x faster than Beta method. This should not come as a surprise, because the implementation of LP-method is remarkably simple (no optimization required; only simple \texttt{Cov} operation of LP-transformed RVs) compared to other kernel based methods which require several level of tuning and pre-processing. Moreover, keep in mind, the other methods are \textit{not automatable} for mixed data problems.

\vskip2.4em
\begin{center}
{\large D. WAIS Data: Shapes of LP-Basis Functions}
\vskip3em
\end{center}
\begin{figure}[h]
\vspace{-1em}
    \centering
    \includegraphics[width=\linewidth,keepaspectratio,trim=.5cm 0cm .5cm .5cm]{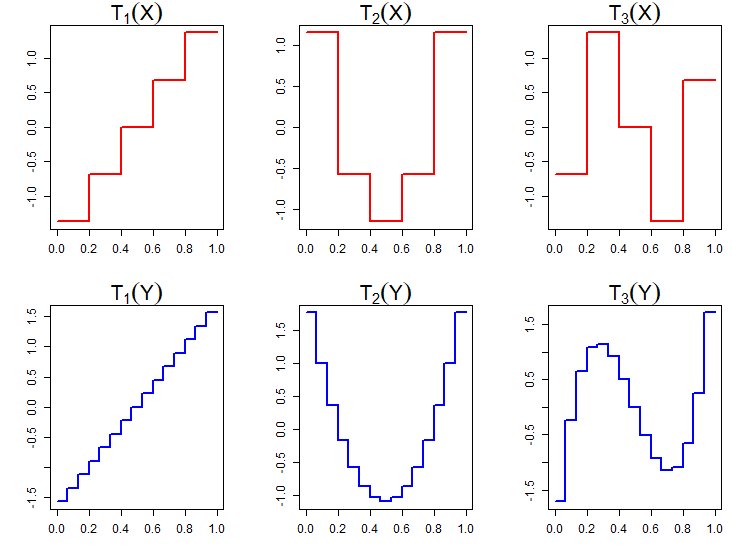}
    \vskip1em
    \caption{WAIS data: The shapes of the top three LP basis functions for the variables X= Age and Y= IQ score. See Sec 6 of the main paper for the description of the data.}
    \label{fig:my_label}
\end{figure}

\end{document}